\documentclass[showkeys,amsmath,amssymb]{revtex4}
\usepackage{bm,amsmath,amsfonts}
\usepackage{color}
\usepackage{listings}            
\usepackage{subfigure}           
\usepackage[urlcolor=blue,linkcolor=blue,colorlinks=true]{hyperref}
\begin{document}
\newcommand{\uu}[1]{\underline{#1}}
\newcommand{\pp}[1]{\phantom{#1}}
\newcommand{\ve}{\varepsilon}
\newcommand{\vs}{\varsigma}
\newcommand{\Tr}{{\,\rm Tr\,}}
\newcommand{\pol}{\frac{1}{2}}
\title{
Relativistic EPR-type experiments for photons in the background of reducible representation HOLA algebras}
\author{Klaudia Wrzask}
\affiliation{
Katedra Fizyki Teoretycznej i Informatyki Kwantowej\\
Politechnika Gda\'nska, 80-233 Gda\'nsk, Poland\\
}
\begin{abstract}
Relativistic EPR-type experiments for photons in the background of reducible representation algebras are discussed. A model for the four Bell states, such that maintains maximally correlated in two bases: linear and circular in all reference frames is developed. Relativistic correlation functions are derived for two cases: when the detectors are transformed under Lorentz transformation in such a way that they still remain in the same reference frame and when just one of the detectors is transformed.
\end{abstract}
\keywords{Bell states, quantum fields, relativistic EPR-type experiments}
\maketitle
\section{Introduction}
Relativistic EPR-type experiments have been discussed at least from 1997 mainly in theoretical background \cite{MC97}-\cite{SZGS02}. This is a good example of a problem where quantum mechanics and theory of relativity are treated under one roof and an occasion to take a closer look at a relativistic model for photon fields in reducible representation of harmonic oscillator Lie algebras (HOLA) proposed by Czachor \cite{MC00}-\cite{MCLecture}.
Some of the difficulties for Lorentz transformation law of Bell states such as the dependence of Wigner rotations on momentum are investigated here in the background of reducible representation. On the other hand, employing relativistic EPR-type experiments in the reducible representation may show the role that the oscillator number $N$ and the vacuum probability density $Z(\bm k)$, known from this representation, play in such model.  \\\\
In \ref{sec:redmot} a basic concept of reducible representations is introduced. Further in \ref{sec:redladder} and \ref{sec:redhola} ladder and number operators are introduced for $N=1$ oscillator space of the theory. How does the extension to $N$-oscillator space look like is discussed in \ref{sec:redladderN}. Finally how to represent vacuum is shown in \ref{sec:redvacuum}.\\\\
Next we develop a model for the four Bell states in the $N$-oscillator reducible representation. First in section \ref{sec:twophotonfieldarbitrary} a two-photon field operator will be shown. Further in \ref{sec:twophotonfieldbell} a model for the Bell state field operators is developed. The main assumption made here, following Zeilinger's paper \cite{AZ97}, is that Bell states are states maximally correlated or maximally anti-correlated in both bases: linear and circular.\\\\
Czachor \cite{MC10} investigated relativistic analogies of EPR states for photons and asked if it is possible to find scalar fields that involve maximal entanglement in two bases in all reference frames. Here we find some answers to this question.  In section \ref{sec:transscalar} we will discuss the effect that the choice of a not unique vacuum has on Lorentz transformations. Further in \ref{sec:transarbitrary} transformation properties of a two-photon state are discussed. In sections \ref{sec:transsinglet} and \ref{sec:transtriplet} the transformation properties of the Bell states are discussed.  \\\\
Further detection in EPR-type experiments in the background of $N$-oscillator reducible representation is discussed. In \ref{sec:detectionEPR} a yes-no observable for describing measurement on detectors is introduced. In \ref{sec:detectiontwophoton} a correlation function for a two-photon state is calculated. In \ref{sec:detectionantycorr} and \ref{sec:detectioncorr} Bell state correlation functions are calculated for maximally correlated and anti-correlated in circular basis states respectively.\\
\\
Finally the transformation rule for the detectors modeled by a yes-no observable is shown and a relativistic correlation function is derived for a two-photon state in the case where two detectors are transformed under Lorentz transformation in such a way that they still maintain in the same reference frame.  Further in \ref{sec:eprtranscase1anty} 
and \ref{sec:eprtranscase1corr} the same calculations are done for maximally anti-correlated and correlated states in circular basis respectively. 
In section \ref{sec:eprtranscase2} a relativistic correlation function is derived for a two-photon state in the case where just one of the detectors is transformed under Lorentz transformation. 
Finally in sections \ref{sec:eprtranscase2anty} and
\ref{sec:eprtranscase2corr} the same calculations are done for maximally anti-correlated and correlated states in circular basis respectively.\\\\
This paper closes with mathematical appendices introducing the ladder operators corresponding to circular and linear polarizations and Lorentz transformation rules corresponding to spin-frames and associated null tetrad.
\section{Reducible representation}
For further analysis let us take a convention $c=\hbar=1$.
When building a relativistic model for photons one has to consider photon's momentum and polarization. In this mathematical model these two quantities will be described in a tensor product structure, i.e.
\begin{eqnarray}
\text{photon momentum space}~
\otimes
~
\text{photon polarization space}.
\end{eqnarray}
It also should be stressed that in relativistic context spin and momentum are not independent degrees of freedom.  Some preliminary aspects of the reducible representation will be discussed here. Such a model has strong arguments in it's favor, mostly because it deals with most of the infrared and ultraviolet divergences. 
~~
\subsection{Motivation}\label{sec:redmot}
In 1925 Heisenberg, Born and Jordan observed that energies of classical free fields look in Fourier space like those of oscillator ensembles. It should be stressed that at that time Heisenberg, Born and Jordan did not know the notation of Fock space and may not fully understood the role of eigenvalues of operators. Having to consider oscillators with different frequencies they considered one oscillator for each frequency mode. The ensemble had to be infinite since the number of modes was infinite.
\\\\
It is a well known problem that standard canonical procedures for field quantization result in various infinity problems.
It was shown in \cite{MC00} by Czachor that the assumption of having one oscillator for each frequency mode may not be natural. This thought continued in a series of papers on reducible representation of CCR \cite{MCMS01} - \cite{MWMC0904}. The main idea for reducible representations is that each of the oscillators is a wave packet, a superposition of infinitely many different momentum states. 
\\
\\
To describe this concept in more detail, let us first introduce a spectral decomposition of a frequency operator
\begin{eqnarray}\label{omega}
\Omega
&=&
\int
d \Gamma (\bm k)~
\omega(\bm k)~
|\bm k\rangle\langle\bm k|,
\end{eqnarray}
so that (\ref{omega}) fulfills the eigenvalue problem
\begin{eqnarray}
\Omega~|\bm k\rangle
&=&
\int
d \Gamma (\bm k')~
\omega(\bm k')~
|\bm k'\rangle\langle\bm k'|
\bm k\rangle
~=~
\omega(\bm k)~
|\bm k\rangle.
\end{eqnarray}
Here $d \Gamma (\bm k)$ in the Lorentz invariant measure 
\begin{equation}\label{Lormeasure}
d\Gamma(\bm{ k})
~=~\frac{d^3k}{(2\pi)^{3} 2 |\bm{k}|}
.
\end{equation}
Furthermore, kets of momentum are normalized to
\begin{equation}
\langle\bm{k}|\bm{k'}\rangle
~=~
(2\pi)^{3}2|\bm{k}|\delta^{(3)}
(\bm{k},\bm{k}')
~=~
\delta_{\Gamma}(\bm{k},\bm{k'}),
\end{equation} 
and the resolution of unity is
\begin{equation}
\int_{R^3}
\frac{d^3 k}{(2\pi)^3 2|\bm k|}
~
|\bm k\rangle \langle \bm k|
~=~
\int_{R^3}
d \Gamma (\bm k)
~
|\bm k\rangle \langle \bm k|
~=~
I.
\end{equation}
The energy for photons, assuming the convention $\hbar=1$, is $E(\bm k)=\omega(\bm k)=|\bm k|$. So the simplest Hamiltonian for one kind of polarization can be written in the form
\begin{eqnarray}
H
&=&
\Omega
\otimes
\left(
a^{\dag}a
+
\frac{1}{2}
\right)
~=~
\int
d \Gamma (\bm k)~
\omega(\bm k)~
|\bm k\rangle\langle\bm k|
\otimes
\left(
a^{\dag}a
+
\frac{1}{2}
\right),
\end{eqnarray}
so that
\begin{eqnarray}
H
|\bm k, n
\rangle
&=&
\omega(\bm k)
\left(
n+\frac{1}{2}
\right)
|\bm k, n
\rangle.
\end{eqnarray}
Here ket $|n\rangle$ is the eigenvector of a ``standard theory" number operator 
$a^{\dagger}a$, where $[a,a^{\dagger}]=1$. 
Now let us introduce an operator that lives in both momentum and polarization spaces
\begin{eqnarray}
a(\bm k,1)
=
|\bm k\rangle\langle\bm k|
\otimes
a.
\end{eqnarray}
Using the resolution of unity, we can also define an operator within the whole spectrum of frequencies
\begin{eqnarray}
a(1)
~=~
\int
d \Gamma (\bm k)~
a(\bm k,1)
&=&
\int
d \Gamma (\bm k)~
|\bm k\rangle\langle\bm k|
\otimes
a
~=~
I
\otimes
a,
\end{eqnarray}
such that the commutator $[a(1),a^{\dagger}(1)]=I\otimes 1$. 
\\
\subsection{Creation and annihilation operators}\label{sec:redladder}
Let us start from the $N=1$ oscillator space with two polarization degrees of freedom. Then the Hilbert space $\mathcal{H}(1)$ of one oscillator is spanned by 
\begin{equation}
|\bm{k},n_{1},n_{2}\rangle
~=~
|\bm{k}\rangle\otimes|n_{1}\rangle
\otimes|n_{2}\rangle.
\end{equation}
Here kets $|n_{\alpha}\rangle$ are eigenvectors of
$a_{\alpha}^{\dagger}a_{\alpha}$. At this point let us consider a two dimensional polarization oscillator and do not determine what kind of polarizations these dimensions determine, only that 
$a_{\alpha}, a_{\alpha'}$
satisfy canonical commutation relations (CCR) typical for irreducible
representation, i.e. 
$[a_{\alpha},a_{\alpha'}^{\dagger}]
=
\delta_{\alpha, \alpha ^{'}}, ~\alpha,\alpha'=1,2$. 
Now the reducible representation of the ladder operators is
\begin{eqnarray}\label{ared1}
a_{\alpha}(\bm{k},1)
&=&
|\bm{k}\rangle\langle\bm{k}|\otimes a_{\alpha},
\\
a_{\alpha}(\bm{k},1)^{\dagger}\label{adagred1}
&=&
|\bm{k}\rangle\langle\bm{k}|\otimes a_{\alpha}^{\dagger}.
\end{eqnarray}
Sub-index $\alpha$ stands for one of the two possible polarization dimensions of an oscillator, where
\begin{equation}
a_{1}~=~
\textrm{a}_1\otimes 1 ,
~~~~~~
a_{2}~=~
1\otimes \textrm{a}_2.
\end{equation}
Then the commutation relations for the reducible representation of the ladder operators are
\begin{equation}
[a_{\alpha}(\bm{k},1),
a_{\alpha'}(\bm{k}',1)^{\dagger}]
~=~
\delta_{\alpha,\alpha'}
\delta_{\Gamma}(\bm{k},\bm{k}')
|\bm{k}\rangle\langle\bm{k}|\otimes 1_2,
\end{equation}
where $1_2$ denotes that it is a tensor product of two $1$. This representation is reducible since the right-hand side of the commutator is an operator valued distribution with 
\begin{eqnarray}
I(\bm{k},1)
~=~|\bm{k}\rangle\langle
\bm{k}|\otimes 1\otimes 1
~=~|\bm{k}\rangle\langle
\bm{k}|\otimes 1_2
\end{eqnarray}
belonging to the center of algebra, i.e. it commutes with the ladder operators
\begin{equation}
[a_{\alpha}(\bm{k},1),I(\bm{k}',1)]
~=~
[a_{\alpha}(\bm{k},1)^{\dagger},I(\bm{k}',1)]
~=~
0.
\end{equation}
Furthermore, the operator $I(\bm{k},1)$ forms the resolution of unity for 
$\mathcal{H}(1)$ Hilbert space:
\begin{equation}
\int d \Gamma(\bm k) ~I(\bm{k},1)
~=~I(1).
\end{equation}
\subsection{HOLA algebra in $N=1$ oscillator space representation}\label{sec:redhola}
The number operator for the reducible representation in $\mathcal{H}(1)$ Hilbert space will be defined as: 
\begin{equation}\label{n1}
n_{\alpha}(\bm{k},1)
~=~
|\bm{k}\rangle\langle\bm{k}|\otimes a_{\alpha}^{\dagger}a_{\alpha}.
\end{equation}
Let us note that this definition is not equivalent to 
\begin{equation}\label{n'1}
n'_{\alpha}(\bm{k},1)
~=~
a_{\alpha}(\bm{k},1)^{\dagger}
a_{\alpha}(\bm{k},1)
~=~
\delta_{\Gamma} (\bm k, \bm k) n_{\alpha}(\bm{k},1).
\end{equation}
We can also define a number operator within the whole spectrum of frequencies as follows
\begin{equation}\label{n(1)}
n_{\alpha}(1)
~=~
a_{\alpha}(1)^{\dag}
a_{\alpha}(1)
~=~
I\otimes a_{\alpha}^{\dagger}a_{\alpha}
~=~
\int 
d\Gamma(\bm k)
|\bm k
\rangle\langle
\bm k
|
\otimes a_{\alpha}^{\dagger}a_{\alpha}
~=~
\int 
d\Gamma(\bm k)
n(\bm k,1).
\end{equation}
So the eigenvalue definition of the number operator, i.e. ``how many photons are there with $\alpha$ polarization within the whole frequency spectrum", would be
\begin{equation}
n_{\alpha}(1)
|\bm{k},n_{1},n_{2}\rangle
~=~
\int d \Gamma(\bm k')
n_{\alpha}(\bm{k}',1)
|\bm{k},n_{1},n_{2}\rangle
~=~
n_{\alpha}
|\bm{k},n_{1},n_{2}\rangle.
\end{equation}
Now the following HOLA for the reducible representation holds:
\begin{eqnarray}
&[a_{\alpha}(\bm{k},1),
a_{\alpha'}(\bm{k}',1)^{\dagger}]
~=&
\delta_{\alpha,\alpha'}
\delta_{\Gamma}(\bm{k},\bm{k}')
I(\bm{k},1),
\\
&[a_{\alpha}(\bm{k},1),
n_{\alpha'}(\bm{k}',1)]
~=&
\delta_{\alpha,\alpha'}
\delta_{\Gamma}(\bm k, \bm k')
a_{\alpha}(\bm{k},1),
\\
&[a_{\alpha}(\bm{k},1)^{\dagger},
n_{\alpha'}(\bm{k}',1)]
~=&
-
\delta_{\alpha,\alpha'}
\delta_{\Gamma}(\bm k, \bm k')
a_{\alpha}(\bm{k},1)^{\dagger} .
\end{eqnarray}
Furthermore, for the representation within the whole frequency spectrum we have:
\begin{eqnarray}
&[a_{\alpha}(1),
a_{\alpha'}(1)^{\dagger}]
~=&
\delta_{\alpha,\alpha'}
I(1),
\\
&[a_{\alpha}(1),
n_{\alpha'}(1)]
~=&
\delta_{\alpha,\alpha'}
a_{\alpha}(1),
\\
&[a_{\alpha}(1)^{\dagger},
n_{\alpha'}(1)]
~=&
-
\delta_{\alpha,\alpha'}
a_{\alpha}(1)^{\dagger} .
\end{eqnarray}
As one can see, the Lie algebra for the whole frequency spectrum holds the ``standard theory" structure.
~~\\
\subsection{HOLA in 
$N$-oscillator space representation}\label{sec:redladderN}
Now let us discuss an extension of the theory to $N$ oscillators. The parameter $N$ characterizes the reducible representation. This parameter is not related to the number of photons.  The Hilbert space in this representation for any $N$ oscillators reads:
\begin{equation}
\mathcal{H}(N)
~=~
\underbrace{\mathcal{H}(1)\otimes\ldots
\otimes\mathcal{H}(1)}_{N}
~=~
\mathcal{H}^{\otimes N},
\end{equation}
and is spanned by kets of the form
\begin{equation}
|\bm{k}_{1},n_{1},n_{2}
\rangle^{(1)}
\otimes
|\bm{k}_{2},n_{1},n_{2}
\rangle^{(2)}
\otimes
\ldots
\otimes
|\bm{k}_{N},n_{1},n_{2}
\rangle^{(N)}.
\end{equation}
Let us also define an operator
\begin{equation}\label{A^(n)}
A^{(n)}
~=~
\underbrace{I\otimes...\otimes I}_{n-1}
\otimes A\otimes\underbrace{I\otimes...\otimes I}_{N-n}.
\end{equation}
The top index $(n)$ shows the ``position" of the $A$ operator in $\mathcal H(N)$ space. 
A natural extension for the ladder operators to the $N$-oscillator space would be
\begin{eqnarray}\label{aN}
a_{\alpha}(\bm{k},N)
&=&
\frac{1}{\sqrt{N}}
\sum_{n=1}^{N}
a_{\alpha}(\bm{k},1)^{(n)}
=
\frac{1}{\sqrt{N}}
\left(
a_{\alpha}(\bm{k},1)
\otimes...\otimes I(1)
+...+
I(1)\otimes...\otimes 
a_{\alpha}(\bm{k},1)
\right).
\end{eqnarray}
The factor $\frac{1}{\sqrt{N}}$ is the normalization factor for the $N$-oscillator representation. 
Let us also define the ladder operators within the whole spectrum of frequencies, i.e.
\begin{eqnarray}\label{a(N)}
a_{\alpha}(N)
&=&
\frac{1}{\sqrt{N}}
\sum_{n=1}^{N}
\int d \Gamma(\bm k )~
a_{\alpha}(\bm{k},1)^{(n)}
=
\frac{1}{\sqrt{N}}
\left(
a_{\alpha}(1)
\otimes...\otimes I(1)
+...+
I(1)\otimes...\otimes 
a_{\alpha}(1)
\right).
\end{eqnarray}
Using the (\ref{n1}) definition of the number operator for $N=1$ representation we write the number operator for the $N$-oscillator representation:
\begin{eqnarray}\label{nIIIN}
n_{\alpha}(\bm{k},N)
&=&
\sum_{n=1}^N \left( |\bm{k}\rangle\langle\bm{k}|
\otimes 
a_{\alpha}^{\dagger}a_{\alpha}
\right)^{(n)}
~=~
\sum_{n=1}^N n_{\alpha}(\bm k, 1)^{(n)}
\nonumber\\
&=&
\left(
n_{\alpha}(\bm{k},1)
\otimes...\otimes I(1)
+...+
I(1)\otimes...\otimes 
n_{\alpha}(\bm{k},1)
\right).
\end{eqnarray}
We also define a number operator within the whole spectrum of frequencies for the $N$-representation
\begin{equation}\label{n(N)}
n_{\alpha}(N)
~=~
\int 
d\Gamma(\bm k)
n(\bm k,N).
\end{equation}
Now let us take a closer look at the Lie algebras for the reducible representation
\begin{eqnarray}
&[a_{\alpha}(\bm{k},N),
a_{\alpha'}(\bm{k}',N)^{\dagger}]\label{CCRN}
~=&
\delta_{\alpha,\alpha'}
\delta_{\Gamma}(\bm{k},\bm{k}')
I(\bm{k},N),
\\
&[a_{\alpha}(\bm{k},N),
n_{\alpha'}(\bm{k}',N)]\label{comanIII}
~=&
\delta_{\alpha,\alpha'}
\delta_{\Gamma}(\bm k, \bm k')
a_{\alpha}(\bm{k},N),
\\
&[a_{\alpha}(\bm{k},N)^{\dag},
n_{\alpha'}(\bm{k}',N)]\label{comadagnIII}
~=&
-
\delta_{\alpha,\alpha'}
\delta_{\Gamma}(\bm k, \bm k')
a_{\alpha}(\bm{k},N)^{\dag}.
\end{eqnarray}
Here at the right-hand side of (\ref{CCRN}) we have an operator:
\begin{equation}
I(\bm{k},N)
~=~
\frac{1}{N}
\sum_{n=1}^N 
I(\bm{k},1)^{(n)}
~=~
\frac{1}{N}
\left(
I(\bm{k},1)
\otimes...\otimes I(1)
+...+
I(1)\otimes...\otimes 
I(\bm{k},1)
\right),
\end{equation}
which for all $N$ is also in the center of algebra since
\begin{equation}
[a_{\alpha}(\bm{k},N),I(\bm{k}',N)]
~=~
[a_{\alpha}(\bm{k},N)^{\dagger},I(\bm{k}',N)]
~=~0.
\end{equation}
Furthermore, the $I(\bm{k},N)$ operator forms the resolution of unity for 
$\mathcal{H}(N)$ Hilbert space:
\begin{eqnarray}
\int d \Gamma(\bm k )~I(\bm{k},N)
~=~I(N)
~=~
\underbrace{I(1)\otimes...\otimes I(1)}_{N}
.
\end{eqnarray}
Again for the $N$-oscillator whole frequency spectrum representation the Lie algebras
\begin{eqnarray}
&[a_{\alpha}(N),
a_{\alpha'}(N)^{\dagger}]
~=&
\delta_{\alpha,\alpha'}
I(N).
\\
&[a_{\alpha}(N),
n_{\alpha'}(N)]\label{}
~=&
\delta_{\alpha,\alpha'}
a_{\alpha}(N),
\\
&[a_{\alpha}(N)^{\dag},
n_{\alpha'}(N)]\label{}
~=&
-
\delta_{\alpha,\alpha'}
a_{\alpha}(N)^{\dag},
\end{eqnarray}
hold the ``standard theory" structure.
\subsection{Vacuum}\label{sec:redvacuum}
The vacuum for $N=1$ oscillator space in reducible representation is any vector of the form
\begin{equation}
|O(1)\rangle~=~\int d\Gamma(\bm{k}) O(\bm{k})
|\bm{k},0,0\rangle.
\end{equation}
This definition implies that vacuum is any vector annihilated by annihilation operators, i.e.
\begin{equation}\label{defvac}
a_{\alpha}(1)|O(1)\rangle~=~0.
\end{equation}
From the normalization condition
\begin{equation}
\langle O(1)
|O(1)\rangle
~=~1
\end{equation}
we get 
\begin{equation}\label{sqrvac}
\int d\Gamma(\bm{k}) |O(\bm{k})|^{2} 
~=~
\int d\Gamma(\bm{k}) Z(\bm{k})
~=~
1.
\end{equation}
Here the scalar field $Z(\bm k)=|O(\bm{k})|^{2} $ represents vacuum probability density. Furthermore, square integrability of (\ref{sqrvac}) implies that $Z(\bm k)$ must decay at infinity and is required to go to zero at $\bm k=0$ in order to avoid infrared divergences \cite{MCJN05}. It should be stressed that this point is of special importance for such reducible representation quantization. It turns out that regularization can be a consequence of employing such special form of scalar field in the definition of vacuum. An  extension to $N$-oscillator space is assumed to be a tensor product of $N=1$ vacuum states, i.e.
\begin{equation}
|O(N)\rangle
~=~
|O(1)\rangle^{\otimes N}
~=~
\underbrace{|O(1)\rangle\otimes...\otimes|O(1)\rangle}_{N}.
\end{equation}
In analogy to (\ref{defvac}) the $N$-oscillator vacuum can be defined as any vector annihilated by $N$-oscillator representation annihilation operators
\begin{equation}
a_{\alpha}(N)|O(N)\rangle~=~0.
\end{equation}
Of course, the normalization condition for the $N$ representation still holds, i.e.
\begin{equation}
\langle O(N)
|O(N)\rangle
~=~
\langle O(1)
|O(1)\rangle^{N}
~=~1.
\end{equation}
\\
\section{Lorentz transformation}
\subsection{Lorentz transformations for the irreducible representation}\label{sec:Lorentzrepresentation}
We start from the irreducible representation and introduce a Lorentz transformation shown by Czachor and Naudts \cite{MCJN05} and further discussed by Czachor and Wrzask \cite{MCKW09}.
The action of such transformation on the annihilation operators is
\begin{eqnarray}
U(\Lambda,\bm{k})^{\dagger}  
a_{1} 
U(\Lambda,\bm{k}) 
&=&
\cos 2\Theta(\Lambda, \bm k) a_{1}
-\sin 2\Theta(\Lambda, \bm k) a_{2}
,
\label{anihilationb1}
\\
U(\Lambda,\bm{k})^{\dagger}
a_{2} 
U(\Lambda,\bm{k})
&=&
\sin 2\Theta(\Lambda, \bm k) a_{1}
+\cos 2\Theta(\Lambda, \bm k) a_{2}
.
\label{anihilationb2}
\end{eqnarray}
Here the unitary operator $U(\Lambda,\bm{k})$ can be written in the form
\begin{eqnarray}
U(\Lambda,\bm{k}) 
&=& 
\exp{\left(-2 i\Theta(\Lambda, \bm k) J_3\right)},\label{UR}
\end{eqnarray}
where $\Theta(\Lambda, \bm k)$ is the Wigner phase, and 
\begin{equation}
J_{3}~=~i(a_{2}^{\dagger}a_{1}
-a_{1}^{\dagger}a_{2})
\end{equation}
is the generator of rotations. $a_1$ and $a_2$ are annihilation operators which correspond to linear polarizations in $x$ and $y$ direction. More on the reducible representation ladder operators corresponding to linear and circular polarizations can be found in appendix.
\subsection{Lorentz transformation for the reducible representation}\label{sec:Lorentzreducible}
To construct a Lorentz transformation for the reducible $N=1$ oscillator
representation the transformation must be written as
\begin{eqnarray}
U(\Lambda,1)\label{LorentzTransN1}\label{WU}
~=~
\int d\Gamma(\bm{k})
|\bm{k}
\rangle\langle 
\bm{\Lambda^{-1}k}|
\otimes
U(\Lambda,\bm k),
\end{eqnarray}
and for the hermitian conjugate we can write
\begin{eqnarray}\label{WUdag}
U(\Lambda,1)^{\dagger}
~=~
\int d\Gamma(\bm{k})
|\bm{\Lambda^{-1}k}
\rangle\langle 
\bm{k}|
\otimes U(\Lambda,\bm k)^{\dag}.
\end{eqnarray}
Furthermore, the $N$-oscillator extension is assumed to be a tensor product of $N=1$ oscillator transformations
\begin{eqnarray}
U(\Lambda,N)\label{LorentzTransN}
&=& 
U(\Lambda,1)^{\otimes N}.
\end{eqnarray}
Then, for circular polarizations under Lorentz transformation we get the following transformation rules for creation and annihilation operators in $N$-oscillator reducible representation.
\begin{eqnarray}
U(\Lambda,N)^{\dagger}
a_s(\bm k,N)
U(\Lambda,N)
&=&
e^{-2is\Theta(\Lambda,\bm{k})}
a_s(\bm{\Lambda^{-1} k},N),
\\
U(\Lambda,N)^{\dagger}
a_s(\bm k,N)^{\dagger}
U(\Lambda,N)
&=&
e^{2is\Theta(\Lambda,\bm{k})}
a_s(\bm{\Lambda^{-1} k},N)^{\dagger},
\\
U(\Lambda,N)
a_s(\bm k,N)
U(\Lambda,N)^{\dagger}
&=&
e^{2is\Theta(\Lambda,\bm{\Lambda k})}
a_s(\bm{\Lambda k},N),
\\
U(\Lambda,N)
a_s(\bm k,N)^{\dagger}
U(\Lambda,N)^{\dagger}
&=&
e^{-2is\Theta(\Lambda,\bm{\Lambda k})}
a_s(\bm{\Lambda k},N)^{\dagger}.
\end{eqnarray}
\subsection{Transformation properties of vacuum}\label{sec:Lorentzvacuum}
Let us remind ourselves that the definition of vacuum in this representation is not unique, i.e.
\begin{equation}
|O(1)\rangle
=
\int d\Gamma(\bm{k}) O(\bm{k})
|\bm{k},0,0\rangle.
\nonumber
\end{equation}
Then the Lorentz transformation acts on vacuum as follows
\begin{eqnarray}\label{vacuumtrans}
U(\Lambda,1)|O(1)\rangle
=
U(\Lambda,1)\int d\Gamma(\bm{k}) O(\bm{k})
|\bm{k},0,0\rangle
=
\int d\Gamma(\bm{k}) O(\bm{\Lambda^{-1}k})
|\bm{k},0,0\rangle
.
\end{eqnarray}
A transformed vacuum state is again a vacuum state, but the probability of finding $\bm k$ is modified by the Doppler effect. The extension to $N > 1$ is obvious. As a by product we observe that the vacuum field transforms as a 
scalar field
\begin{eqnarray}
O(\bm{k})
~~\mapsto~~
O(\bm{\Lambda^{-1}k})
.
\end{eqnarray}
This also implies the following transformation rule of the vacuum probability density
\begin{eqnarray}
Z(\bm{k})
~~\mapsto~~
Z(\bm{\Lambda^{-1}k})
.
\end{eqnarray}
Of course, the norm of such a transformed vacuum is invariant due to the Lorentz invariant measure (\ref{Lormeasure}), and therefore
\begin{equation}
\langle O,1| U(\Lambda,1)^{\dagger} U(\Lambda,1)
|O(1)\rangle ~=~
\int d\Gamma(\bm{k}) |O(\bm{\Lambda^{-1}k})|^{2} ~=~
\int d\Gamma(\bm{\Lambda k}) |O(\bm{k})|^{2} ~=~1.
\end{equation}
\\
\section{Two-photon field}\label{sec:twophotonfield}
\subsection{Two-photon field operator}\label{sec:twophotonfieldarbitrary}
Let us first note that the correlation of the two-photon states in the $N$-oscillator reducible representation doesn't come straight forward form the tensor product. To see this, let us first rewrite the ladder operators in the $N$-oscillator representation, i.e.
\begin{eqnarray}
a_s(\bm k,N)^{\dagger}\label{tensoraa}
a_{s'}(\bm k',N)^{\dagger}
&=&
\frac{1}{\sqrt{N}}
\sum_{n}^{N}
a_{s}(\bm{k},1)^{\dagger (n)}
\frac{1}{\sqrt{N}}
\sum_{m}^{N}
a_{s'}(\bm{k}',1)^{\dagger (m)}
\nonumber\\
&=&
\frac{1}{N}
\sum_{n=m}^{N}
a_{s}(\bm{k},1)^{\dagger (n)}
a_{s'}(\bm{k}',1)^{\dagger (m)}
+
\frac{1}{N}
\sum_{n\ne m}^{N}
a_{s}(\bm{k},1)^{\dagger (n)}
a_{s'}(\bm{k}',1)^{\dagger (m)}
\nonumber\\
&=&
\frac{1}{N}
\underbrace{
\left(
a_{s}(\bm{k},1)^{\dagger}
a_{s'}(\bm{k}',1)^{\dagger}
\otimes...\otimes I(1)
+...+
I(1)\otimes...\otimes 
a_{s}(\bm{k},1)^{\dagger}
a_{s'}(\bm{k}',1)^{\dagger}
\right)}_{N}
\nonumber\\
&+&
\frac{1}{N}
\underbrace{
\left(
a_{s}(\bm{k},1)^{\dagger}
\otimes a_{s'}(\bm{k}',1)^{\dagger}
\otimes...\otimes I(1)
+...+
I(1)\otimes...\otimes 
a_{s}(\bm{k},1)^{\dagger}
\otimes a_{s'}(\bm{k}',1)^{\dagger}
\right)}_{N^2-N}.
\nonumber\\
\end{eqnarray}
As one can see, in (\ref{tensoraa}) we have $N$ terms where the creation operators live on the same oscillator and $N^2-N$ terms where they live on separate oscillators.  Being in the subject of tensor algebra representations, it is worth mentioning that the choice the algebra representation may become important. For example, it was shown by Paw\l{}owski and Czachor \cite{MPMC06} that the ``entanglement with vacuum" turns out to be algebra representation dependent.\\
\\
Now let us consider a whole frequency spectrum two-photon field operator in circular basis for the $N$-oscillator reducible representation
\begin{eqnarray}\label{PsiNs}
\Psi(N)
&=&
\sum_{s,s'=\pm}
\int d \Gamma(\bm k)  d\Gamma(\bm{k}') ~ 
\psi_{ss'}(\bm k,\bm k')
a_s(\bm k,N)^{\dagger}
a_{s'}(\bm k',N)^{\dagger}.
\end{eqnarray}
To study more the symmetry properties of such two-photon operators we write:
\begin{eqnarray}\label{PsiNssym}
\Psi(N)
&=&
\sum_{s,s'=\pm}
\int d\Gamma(\bm k)  d\Gamma(\bm{k}') ~ 
\frac{
\psi_{ss'}(\bm k,\bm k')+
\psi_{s's}(\bm k',\bm k)
}{2}
a_s(\bm k,N)^{\dagger}
a_{s'}(\bm k',N)^{\dagger}.
\end{eqnarray}
Let us also note that, from the integral's (\ref{PsiNssym}) symmetry property, it follows 
\begin{eqnarray}\label{symcond}
\psi_{ss'}(\bm k,\bm k')
=
\psi_{s's}(\bm k',\bm k).
\end{eqnarray}
From the relation between the circular and linear polarizations derived in the appendix (\ref{astotheta}), we can get to the linear polarization basis:
\begin{eqnarray}\label{PsiNtheta}
\Psi(N)
&=&
\sum_{s,s'}
\int d\Gamma(\bm k)  d\Gamma(\bm{k}') ~ 
e^{i (s\theta(\bm k)+s'\theta(\bm k')) }
\frac{\psi_{ss'}(\bm k,\bm k')
}{2}
\left(
a_{\theta}(\bm k,N)^{\dagger}
a_{\theta}(\bm k',N)^{\dagger}
-
ss' 
a_{\theta'}(\bm k,N)^{\dagger}
a_{\theta'}(\bm k',N)^{\dagger}
\right)
\nonumber\\
&+&
\sum_{s,s'}
\int d\Gamma(\bm k)  d\Gamma(\bm{k}') ~ 
e^{i (s\theta(\bm k)+s'\theta(\bm k'))}
\frac{\psi_{ss'}(\bm k,\bm k')
}{2}
\left(
i s' 
a_{\theta}(\bm k,N)^{\dagger}
a_{\theta'}(\bm k',N)^{\dagger}
+
i s 
a_{\theta'}(\bm k,N)^{\dagger}
a_{\theta}(\bm k',N)^{\dagger}
\right)
\nonumber\\
\\
&=&
\sum_{s,s'}
\int d\Gamma(\bm k)  d\Gamma(\bm{k}') ~ 
e^{i (s\theta(\bm k)+s'\theta(\bm k'))}
\frac{
\psi_{ss'}(\bm k,\bm k')
}{2}
\left(
a_{\theta}(\bm k,N)^{\dagger}
a_{\theta}(\bm k',N)^{\dagger}
-
ss' 
a_{\theta'}(\bm k,N)^{\dagger}
a_{\theta'}(\bm k',N)^{\dagger}
\right)
\nonumber\\
&+&i
\sum_{s,s'}
\int d\Gamma(\bm k)  d\Gamma(\bm{k}') ~ 
s'
e^{i (s\theta(\bm k)+s'\theta(\bm k'))}
\psi_{ss'}(\bm k,\bm k')
a_{\theta}(\bm k,N)^{\dagger}
a_{\theta'}(\bm k',N)^{\dagger}.
\end{eqnarray}
Then the scalar product of such two-photon state reads
\begin{eqnarray}\label{scalarPsiPsi}
&&
\langle O(N)|
\Psi(N)^{\dagger}
\Psi(N)
|O(N)\rangle
\hspace{11.3 cm}
\nonumber\\
&=&
\frac{2}{N}
\sum_{s,s'=\pm}
\int d\Gamma(\bm k) 
|\psi_{ss'}(\bm k,\bm k)|^2
Z(\bm k)
+
\frac{2(N-1)}{N}
\sum_{s,s'=\pm}
\int d\Gamma(\bm k)  d\Gamma(\bm{k}')
|\psi_{ss'}(\bm k,\bm k')|^2
Z(\bm k)Z(\bm k').
\end{eqnarray}
Also in the $N\to\infty$ limit we get
\begin{eqnarray}\label{scalarPsiPsiNinfty}
&&
\lim_{N\to\infty}
\langle O(N)|
\Psi(N)^{\dagger}
\Psi(N)
|O(N)\rangle
~=~
2
\sum_{s,s'=\pm}
\int d\Gamma(\bm k)  d\Gamma(\bm{k}')
|\psi_{ss'}(\bm k,\bm k')|^2
Z(\bm k)Z(\bm k').
\end{eqnarray}
As one can see, the first term of the scalar product (\ref{scalarPsiPsi}) 
does not occur in $N\to\infty$ limit.
\\
\subsection{Bell state field operators}\label{sec:twophotonfieldbell}
Now let us study some cases of maximal photon correlations in a quantum field theory background for the reducible $N$-oscillator representation.
First we will consider photons in circular basis that are anti-correlated: one is left- and the other right-handed. Anti-correlated in circular basis field operators will be denoted by $\Psi_1(N)$, so that
\begin{eqnarray}\label{Psi1}
\Psi_1(N)
&=&
\sum_{s\ne s'}
\int d\Gamma(\bm k)  d\Gamma(\bm{k}') ~ 
\psi_{ss'}(\bm k,\bm k')
a_s(\bm k,N)^{\dagger}
a_{s'}(\bm k',N)^{\dagger}
\nonumber\\
&=&
\int d\Gamma(\bm k)  d\Gamma(\bm{k}') ~ 
\left(
\psi_{+-}(\bm k,\bm k')
a_{+}(\bm k,N)^{\dagger}
a_{-}(\bm k',N)^{\dagger}
+
\psi_{-+}(\bm k,\bm k')
a_{-}(\bm k,N)^{\dagger}
a_{+}(\bm k',N)^{\dagger}
\right)\label{Psi1cir}
\\
&=&
\sum_{s\ne s'}
\int d\Gamma(\bm k)  d\Gamma(\bm{k}') ~ 
e^{i (s\theta(\bm k)+s'\theta(\bm k'))}
\frac{\psi_{ss'}(\bm k,\bm k')
}{2}
\left(
a_{\theta}(\bm k,N)^{\dagger}
a_{\theta}(\bm k',N)^{\dagger}
+
a_{\theta'}(\bm k,N)^{\dagger}
a_{\theta'}(\bm k',N)^{\dagger}
\right)
\nonumber\\
&+&i
\sum_{s\ne s'}
\int d\Gamma(\bm k)  d\Gamma(\bm{k}') ~ 
s'
e^{i (s\theta(\bm k)+s'\theta(\bm k'))}
\psi_{ss'}(\bm k,\bm k')
a_{\theta}(\bm k,N)^{\dagger}
a_{\theta'}(\bm k',N)^{\dagger}.\label{Psi1lin}
\end{eqnarray}
We can say about $\Psi_1(N)$: both photons have different polarizations in circular basis. For the maximal anti-correlation the condition on the field must hold $|\psi_{+-}(\bm k,\bm k')|^2=|\psi_{-+}(\bm k,\bm k')|^2$. 
To fully describe four Bell states we need a second basis. This is why we write the $\Psi_1(N)$ field operator in two bases: circular (\ref{Psi1cir}) and linear (\ref{Psi1lin}). Let us stress that for the linear basis the polarization angle $\theta(\bm k)$ is assumed to be dependent on momentum. 
There are two situations when such states are still maximally correlated in the second, here linear basis. The first case is when photons in linear basis are maximally anti-correlated. Let us denote $\theta_{11}(\bm k)$ as the polarization function for such states. Then, from (\ref{Psi1lin}), the following condition on the field and polarization angle must hold:
\begin{eqnarray}\label{bellcond11}
e^{i(\theta_{11}(\bm k)-\theta_{11}(\bm k'))}
\psi_{+-}(\bm k,\bm k')
=
- e^{i(\theta_{11}(\bm k')-\theta_{11}(\bm k))}
\psi_{-+}(\bm k,\bm k'),
\end{eqnarray}
and such field operator, denoted here by $\Psi_{11}(N)$, may be written in forms
\begin{eqnarray}\label{Psi11}
\Psi_{11}(N)
&=&
\int d\Gamma(\bm k)  d\Gamma(\bm{k}') ~ 
\left(
\psi_{+-}(\bm k,\bm k')
a_{+}(\bm k,N)^{\dagger}
a_{-}(\bm k',N)^{\dagger}
+
\psi_{-+}(\bm k,\bm k')
a_{-}(\bm k,N)^{\dagger}
a_{+}(\bm k',N)^{\dagger}
\right)
\nonumber\\
&=&
-i
\sum_{s=\pm}
\int d\Gamma(\bm k)  d\Gamma(\bm{k}') ~ 
s
e^{i (s\theta_{11}(\bm k)-s\theta_{11}(\bm k'))}
\psi_{s-s}(\bm k,\bm k')
a_{\theta}(\bm k,N)^{\dagger}
a_{\theta'}(\bm k',N)^{\dagger}
\nonumber\\
&=&
-2i
\int d\Gamma(\bm k)  d\Gamma(\bm{k}') ~ 
e^{i (\theta_{11}(\bm k)-\theta_{11}(\bm k'))}
\psi_{+-}(\bm k,\bm k')
a_{\theta}(\bm k,N)^{\dagger}
a_{\theta'}(\bm k',N)^{\dagger}
\nonumber\\
&=&
2i
\int d\Gamma(\bm k)  d\Gamma(\bm{k}') ~ 
e^{-i (\theta_{11}(\bm k)-\theta_{11}(\bm k'))}
\psi_{-+}(\bm k,\bm k')
a_{\theta}(\bm k,N)^{\dagger}
a_{\theta'}(\bm k',N)^{\dagger}
.
\end{eqnarray}
From (\ref{bellcond11}) we see that if we want to have a two-photon state maximally correlated in both bases, an implicit relation must take place that relates the fields $\psi_{+-}(\bm k,\bm k')$ with the polarization angles. Now we can say about  $\Psi_{11}(N)$: both photons have different polarizations in circular and linear basis. In such a case operator (\ref{Psi11}) will represent one of the four Bell states.
Then the inner product reads
\begin{eqnarray}\label{scalarpsi11psi11}
&&
\langle O(N)|
\Psi_{11}(N)^{\dagger}
\Psi_{11}(N)
|O(N)\rangle
\nonumber\\
&=&
\frac{4}{N}
\int d\Gamma(\bm k) 
|\psi_{+-}(\bm k,\bm k)|^2
Z(\bm k)
+
\frac{4(N-1)}{N}
\int d\Gamma(\bm k)  d\Gamma(\bm{k}')
|\psi_{+-}(\bm k,\bm k')|^2
Z(\bm k)Z(\bm k').
\end{eqnarray}
The second case is when photons are maximally correlated in linear basis. Then from (\ref{Psi1lin}) a condition on the fields and the polarization angles must hold:
\begin{eqnarray}\label{bellcond12}
e^{i(\theta_{12}(\bm k)-\theta_{12}(\bm k'))}
\psi_{+-}(\bm k,\bm k')
&=&
e^{-i(\theta_{12}(\bm k)-\theta_{12}(\bm k'))}
\psi_{-+}(\bm k,\bm k').
\end{eqnarray}
Here we denote $\theta_{12}(\bm k)$ as the polarization function for such a field operator 
and $\Psi_{12}(N)$ as the field operator corresponding to the Bell state that is anti-correlated in circular basis and correlated in the linear one, i.e.
\begin{eqnarray}\label{Psi12}
\Psi_{12}(N)
&=&
\int d\Gamma(\bm k)  d\Gamma(\bm{k}') ~ 
\left(
\psi_{+-}(\bm k,\bm k')
a_{+}(\bm k,N)^{\dagger}
a_{-}(\bm k',N)^{\dagger}
+
\psi_{-+}(\bm k,\bm k')
a_{-}(\bm k,N)^{\dagger}
a_{+}(\bm k',N)^{\dagger}
\right)
\nonumber\\
&=&
\int d\Gamma(\bm k)  d\Gamma(\bm{k}') ~ 
e^{i (\theta_{12}(\bm k)-\theta_{12}(\bm k'))}
\psi_{+-}(\bm k,\bm k')
\left(
a_{\theta}(\bm k,N)^{\dagger}
a_{\theta}(\bm k',N)^{\dagger}
+
a_{\theta'}(\bm k,N)^{\dagger}
a_{\theta'}(\bm k',N)^{\dagger}
\right)
\nonumber\\
&=&
\int d\Gamma(\bm k)  d\Gamma(\bm{k}') ~ 
e^{-i (\theta_{12}(\bm k)-\theta_{12}(\bm k'))}
\psi_{-+}(\bm k,\bm k')
\left(
a_{\theta}(\bm k,N)^{\dagger}
a_{\theta}(\bm k',N)^{\dagger}
+
a_{\theta'}(\bm k,N)^{\dagger}
a_{\theta'}(\bm k',N)^{\dagger}
\right).
\nonumber\\
\end{eqnarray}
Also from conditions (\ref{bellcond11}) and (\ref{bellcond12}) we get the following relation for the polarization angles
\begin{eqnarray}\label{rel11-12}
\theta_{11}(\bm k)-\theta_{11}(\bm k')
~=~
\theta_{12}(\bm k)-\theta_{12}(\bm k')+\frac{\pi}{2}+n\pi,~~~~n\in\bm{Z}.
\end{eqnarray}
This formula seems to be more intuitive than conditions (\ref{bellcond11}) and (\ref{bellcond12}). It simply shows that the difference of the polarization angles for anti-correlated linear polarizations states is equal to the difference of the polarization angles for correlated states plus a $\pi/2$ factor.
Now we can say about $\Psi_{12}(N)$: both photons have different polarizations in circular basis and the same polarizations in the linear one. The inner product for (\ref{Psi12}) state reads
\begin{eqnarray}\label{scalarpsi12psi12}
&&
\langle O(N)|
\Psi_{12}(N)^{\dagger}
\Psi_{12}(N)
|O(N)\rangle
\nonumber\\
&=&
\frac{4}{N}
\int d\Gamma(\bm k) 
|\psi_{+-}(\bm k,\bm k)|^2
Z(\bm k)
+
\frac{4(N-1)}{N}
\int d\Gamma(\bm k)  d\Gamma(\bm{k}')
|\psi_{+-}(\bm k,\bm k')|^2
Z(\bm k)Z(\bm k').
\end{eqnarray}
Now let us consider photons in circular basis that are maximally correlated, i.e.  $|\psi_{++}(\bm k,\bm k')|^2=|\psi_{--}(\bm k,\bm k')|^2$; they're both either left or right-handed. We will denote such field operators as $\Psi_2(N)$ and write them in both bases
\begin{eqnarray}\label{Psi2}
\Psi_2(N)
&=&
\sum_{s= s'}
\int d\Gamma(\bm k)  d\Gamma(\bm{k}') ~ 
\psi_{ss'}(\bm k,\bm k')
a_s(\bm k,N)^{\dagger}
a_{s'}(\bm k',N)^{\dagger}
\nonumber\\
&=&
\int d\Gamma(\bm k)  d\Gamma(\bm{k}') ~ 
\left(
\psi_{++}(\bm k,\bm k')
a_{+}(\bm k,N)^{\dagger}
a_{+}(\bm k',N)^{\dagger}
+
\psi_{--}(\bm k,\bm k')
a_{-}(\bm k,N)^{\dagger}
a_{-}(\bm k',N)^{\dagger}
\right)
\\
&=&
\sum_{s= s'}
\int d\Gamma(\bm k)  d\Gamma(\bm{k}') ~ 
e^{i (s\theta(\bm k)+s'\theta(\bm k'))}
\frac{\psi_{ss'}(\bm k,\bm k')
}{2}
\left(
a_{\theta}(\bm k,N)^{\dagger}
a_{\theta}(\bm k',N)^{\dagger}
-
a_{\theta'}(\bm k,N)^{\dagger}
a_{\theta'}(\bm k',N)^{\dagger}
\right)
\nonumber\\
&+&i
\sum_{s= s'}
\int d\Gamma(\bm k)  d\Gamma(\bm{k}') ~ 
s'
e^{i (s\theta(\bm k)+s'\theta(\bm k'))}
\psi_{ss'}(\bm k,\bm k')
a_{\theta}(\bm k,N)^{\dagger}
a_{\theta'}(\bm k',N)^{\dagger}.
\end{eqnarray}
Again there are two situations when such a field operator is still maximally correlated in another, here linear basis. When such photons in linear basis are maximally anti-correlated, a condition on fields $\psi_{--}(\bm k,\bm k')$ and $\psi_{++}(\bm k,\bm k')$ and the polarization angle denoted here by $\theta_{21}(\bm k)$ must hold
\begin{eqnarray}\label{bellcond21}
e^{i (\theta_{21}(\bm k)+\theta_{21}(\bm k'))}
\psi_{++}(\bm k,\bm k')
&=&
-
e^{-i (\theta_{21}(\bm k)+\theta_{21}(\bm k'))}
\psi_{--}(\bm k,\bm k').
\end{eqnarray}
Then the field operator $\Psi_{21}(N)$ can be written in the following forms
\begin{eqnarray}\label{Psi21}
\Psi_{21}(N)
&=&
\int d\Gamma(\bm k)  d\Gamma(\bm{k}') ~ 
\left(
\psi_{++}(\bm k,\bm k')
a_{+}(\bm k,N)^{\dagger}
a_{+}(\bm k',N)^{\dagger}
+
\psi_{--}(\bm k,\bm k')
a_{-}(\bm k,N)^{\dagger}
a_{-}(\bm k',N)^{\dagger}
\right)
\nonumber\\
&=&i
\sum_{s= \pm}
\int d\Gamma(\bm k)  d\Gamma(\bm{k}') ~ 
s
e^{i (s\theta_{21}(\bm k)+s\theta_{21}(\bm k'))}
\psi_{ss}(\bm k,\bm k')
a_{\theta}(\bm k,N)^{\dagger}
a_{\theta'}(\bm k',N)^{\dagger}
\nonumber\\
&=&
2i
\int d\Gamma(\bm k)  d\Gamma(\bm{k}') ~ 
e^{i (\theta_{21}(\bm k)+\theta_{21}(\bm k'))}
\psi_{++}(\bm k,\bm k')
a_{\theta}(\bm k,N)^{\dagger}
a_{\theta'}(\bm k',N)^{\dagger}
\nonumber\\
&=&
-2i
\int d\Gamma(\bm k)  d\Gamma(\bm{k}') ~ 
e^{-i (\theta_{21}(\bm k)+\theta_{21}(\bm k'))}
\psi_{--}(\bm k,\bm k')
a_{\theta}(\bm k,N)^{\dagger}
a_{\theta'}(\bm k',N)^{\dagger}
.
\end{eqnarray}
We can say about $\Psi_{21}(N)$: both photons have the same polarizations in circular and different polarizations in linear basis. Then the inner product reads
\begin{eqnarray}\label{scalarpsi21psi21}
&&
\langle O(N)|
\Psi_{21}(N)^{\dagger}
\Psi_{21}(N)
|O(N)\rangle
\nonumber\\
&=&
\frac{4}{N}
\int d\Gamma(\bm k) 
|\psi_{++}(\bm k,\bm k)|^2
Z(\bm k)
+
\frac{4(N-1)}{N}
\int d\Gamma(\bm k)  d\Gamma(\bm{k}')
|\psi_{++}(\bm k,\bm k')|^2
Z(\bm k)Z(\bm k').
\end{eqnarray}
Finally when photons in linear basis are maximally correlated, a condition on the fields and polarization angles, denoted here by $\theta_{22}(\bm k)$, must hold
\begin{eqnarray}\label{bellcond22}
e^{i (\theta_{22}(\bm k)+\theta_{22}(\bm k'))}
\psi_{++}(\bm k,\bm k')
&=&
e^{-i (\theta_{22}(\bm k)+\theta_{22}(\bm k'))}
\psi_{--}(\bm k,\bm k'),
\end{eqnarray}
and the field operator denoted here by $\Psi_{22}(N)$ is
\begin{eqnarray}\label{Psi22}
\Psi_{22}(N)
&=&
\int d\Gamma(\bm k)  d\Gamma(\bm{k}') ~ 
\left(
\psi_{++}(\bm k,\bm k')
a_{+}(\bm k,N)^{\dagger}
a_{+}(\bm k',N)^{\dagger}
+
\psi_{--}(\bm k,\bm k')
a_{-}(\bm k,N)^{\dagger}
a_{-}(\bm k',N)^{\dagger}
\right)
\nonumber\\
&=&
\sum_{s= \pm}
\int d\Gamma(\bm k)  d\Gamma(\bm{k}') ~ 
e^{i (s\theta_{22}(\bm k)+s\theta_{22}(\bm k'))}
\frac{\psi_{ss}(\bm k,\bm k')
}{2}
\left(
a_{\theta}(\bm k,N)^{\dagger}
a_{\theta}(\bm k',N)^{\dagger}
-
a_{\theta'}(\bm k,N)^{\dagger}
a_{\theta'}(\bm k',N)^{\dagger}
\right)
\nonumber\\
&=&
\int d\Gamma(\bm k)  d\Gamma(\bm{k}') ~ 
e^{i (\theta_{22}(\bm k)+\theta_{22}(\bm k'))}
\psi_{++}(\bm k,\bm k')
\left(
a_{\theta}(\bm k,N)^{\dagger}
a_{\theta}(\bm k',N)^{\dagger}
-
a_{\theta'}(\bm k,N)^{\dagger}
a_{\theta'}(\bm k',N)^{\dagger}
\right)
\nonumber\\
&=&
\int d\Gamma(\bm k)  d\Gamma(\bm{k}') ~ 
e^{-i (\theta_{22}(\bm k)+\theta_{22}(\bm k'))}
\psi_{--}(\bm k,\bm k')
\left(
a_{\theta}(\bm k,N)^{\dagger}
a_{\theta}(\bm k',N)^{\dagger}
-
a_{\theta'}(\bm k,N)^{\dagger}
a_{\theta'}(\bm k',N)^{\dagger}
\right).
\nonumber
\\
\end{eqnarray}
Also from conditions (\ref{bellcond21}) and (\ref{bellcond22}) we get the following relation for the polarization angles
\begin{eqnarray}\label{rel21-22}
\theta_{21}(\bm k)+\theta_{21}(\bm k')
~=~
\theta_{22}(\bm k)+\theta_{22}(\bm k')+\frac{\pi}{2}+n\pi,~~~~n\in\bm{Z}.
\end{eqnarray}
We can say about $\Psi_{22}(N)$: both photons have the same polarizations in circular and  linear basis. Then the inner product reads
\begin{eqnarray}\label{scalarpsi22psi22}
&&
\langle O(N)|
\Psi_{22}(N)^{\dagger}
\Psi_{22}(N)
|O(N)\rangle
\nonumber\\
&=&
\frac{4}{N}
\int d\Gamma(\bm k) 
|\psi_{++}(\bm k,\bm k)|^2
Z(\bm k)
+
\frac{4(N-1)}{N}
\int d\Gamma(\bm k)  d\Gamma(\bm{k}')
|\psi_{++}(\bm k,\bm k')|^2
Z(\bm k)Z(\bm k').
\end{eqnarray}
%
\\
We will refer to operators (\ref{Psi11}), (\ref{Psi12}), (\ref{Psi21}), (\ref{Psi22}) as the four Bell state corresponding field operators.\\
\section{Transformation properties of two-photon fields}\label{sec:trans}
\subsection{Scalar field}\label{sec:transscalar}
First let us consider an invariant two-photon field operator $\Psi(N)$, i.e.
\begin{eqnarray}
U(\Lambda,N)
\Psi(N)
U(\Lambda,N)^{\dagger}
&=&
\Psi(N).
\end{eqnarray}
In reducible representation, where vacuum is not unique, performing a Lorentz transformation on states corresponding to invariant field operators doesn't result in the same state since the vacuum also transforms as a scalar field (\ref{vacuumtrans}), so that
\begin{eqnarray}
U(\Lambda,N)
|\Psi(N)\rangle
~=~
U(\Lambda,N)
\Psi(N)
U(\Lambda,N)^{\dagger}
U(\Lambda,N)
|O(N)\rangle
~=~
\Psi(N)
|O_{\Lambda}(N)\rangle.
\end{eqnarray}
Of course, we make the assumption that the scalar product is conserved under Lorentz transformation, so that
\begin{eqnarray}
&&
\langle O_{\Lambda}(N)|
\Psi(N)^{\dagger}
\Psi(N)
|O_{\Lambda}(N)\rangle
\nonumber\\
&=&
2
\sum_{s,s'=\pm}
\int d\Gamma(\bm k)  d\Gamma(\bm{k}')
|\psi_{ss'}(\bm k,\bm k')|^2
\langle O_{\Lambda}(N)|
I(\bm k,N)I(\bm k',N)
|O_{\Lambda}(N)\rangle
\nonumber\\
&=&
\frac{2}{N}
\sum_{s,s'=\pm}
\int d\Gamma(\bm k) 
|\psi_{ss'}(\bm k,\bm k)|^2
Z(\bm{\Lambda^{-1} k})
+
\frac{2(N-1)}{N}
\sum_{s,s'=\pm}
\int d\Gamma(\bm k)  d\Gamma(\bm{k}')
|\psi_{ss'}(\bm k,\bm k')|^2
Z(\bm{\Lambda^{-1} k})Z(\bm{\Lambda^{-1} k}')
\nonumber\\
&=&
\frac{2}{N}
\sum_{s,s'=\pm}
\int d\Gamma(\bm k) 
|\psi_{ss'}(\bm{\Lambda k},\bm{\Lambda k})|^2
Z(\bm{ k})
+
\frac{2(N-1)}{N}
\sum_{s,s'=\pm}
\int d\Gamma(\bm k)  d\Gamma(\bm{k}')
|\psi_{ss'}(\bm{\Lambda k},\bm{\Lambda k}')|^2
Z(\bm{ k})Z(\bm{ k}'),
\nonumber\\
\end{eqnarray}
and this implies the following condition on the field
\begin{eqnarray}\label{condtransfield}
|\psi_{ss'}(\bm{\Lambda k},\bm{\Lambda k}')|^2
&=&
|\psi_{ss'}(\bm{k},\bm{k}')|^2.
\end{eqnarray}
\subsection{Transformation properties of a two-photon field operator}\label{sec:transarbitrary}
First let us state that under Lorentz transformation we have the following transformation rules for creation operators in circular basis:
\begin{equation}
U(\Lambda,N)^{\dagger}
a_s(\bm k,N)^{\dagger}
a_{s'}(\bm k',N)^{\dagger}
U(\Lambda,N)
~=~
e^{2is\Theta(\Lambda,\bm{k})}
e^{2is'\Theta(\Lambda,\bm{k}')}
a_s(\bm{\Lambda^{-1} k},N)^{\dagger}
a_{s'}(\bm{\Lambda^{-1} k}',N)^{\dagger},
\end{equation}
\begin{equation}
U(\Lambda,N)
a_s(\bm k,N)^{\dagger}
a_{s'}(\bm k',N)^{\dagger}
U(\Lambda,N)^{\dagger}
~=~
e^{-2is\Theta(\Lambda,\bm{\Lambda k})}
e^{-2is'\Theta(\Lambda,\bm{\Lambda k}')}
a_s(\bm{\Lambda k},N)^{\dagger}
a_{s'}(\bm{\Lambda k}',N)^{\dagger}.
\end{equation}
Now we will consider a two-photon field operator in circular basis. Let us assume that the field operator $\Psi(N)$ (\ref{PsiNs}) satisfies the scalar field condition, so that under Lorentz transformation we have
\begin{eqnarray}
&&
U(\Lambda,N)
\Psi(N)
U(\Lambda,N)^{\dagger}
\nonumber\\
&=&
U(\Lambda,N)
\sum_{s,s'}
\int d\Gamma(\bm k)  d\Gamma(\bm{k}') ~ 
\psi_{ss'}(\bm k,\bm k')
a_s(\bm k,N)^{\dagger}
a_{s'}(\bm k',N)^{\dagger}
U(\Lambda,N)^{\dagger}
\nonumber\\
&=&
\sum_{s,s'}
\int d\Gamma(\bm k)  d\Gamma(\bm{k}') ~ 
\psi_{ss'}(\bm k,\bm k')
e^{-2is\Theta(\Lambda,\bm{\Lambda k})}
e^{-2is'\Theta(\Lambda,\bm{\Lambda k}')}
a_s(\bm{\Lambda k},N)^{\dagger}
a_{s'}(\bm{\Lambda k}',N)^{\dagger}
\nonumber\\
&=&
\sum_{s,s'}
\int d\Gamma(\bm k)  d\Gamma(\bm{k}') ~ 
\psi_{ss'}(\bm{\Lambda^{-1} k},\bm{\Lambda^{-1} k}')
e^{-2is\Theta(\Lambda,\bm{k})}
e^{-2is'\Theta(\Lambda,\bm{k}')}
a_s(\bm{k},N)^{\dagger}
a_{s'}(\bm{k}',N)^{\dagger}
\nonumber\\
&=&
\Psi(N).
\end{eqnarray}
This implies the following transformation rule for the fields:
\begin{eqnarray}
\psi_{ss'}(\bm k,\bm k')\label{Psiss'trans}
e^{2is\Theta(\Lambda,\bm{k})}
e^{2is'\Theta(\Lambda,\bm{k}')}
&=&
\psi_{ss'}(\bm{\Lambda^{-1} k},\bm{ \Lambda^{-1} k}'),
\end{eqnarray}
which is consistent with (\ref{condtransfield}).
\subsection{Transformation properties of states maximally anti-correlated in circular basis}\label{sec:transsinglet}
Now let us assume that the field operator $\Psi_{11}(N)$ corresponding to one of the Bell states (\ref{Psi11}) transforms as a scalar field under Lorentz transformation, so that
\begin{eqnarray}
&&
U(\Lambda,N)
\Psi_{11}(N)
U(\Lambda,N)^{\dagger}
\nonumber\\
&=&
U(\Lambda,N)
\sum_{s=\pm}
\int d\Gamma(\bm k)  d\Gamma(\bm{k}') ~ 
\psi_{s-s}(\bm k,\bm k')
a_{s}(\bm k,N)^{\dagger}
a_{-s}(\bm k',N)^{\dagger}
U(\Lambda,N)^{\dagger}
\nonumber\\
&=&
\int d\Gamma(\bm k)  d\Gamma(\bm{k}') ~ 
\psi_{+-}(\bm k,\bm k')
e^{-2i\Theta(\Lambda,\bm{\Lambda k})}
e^{2i\Theta(\Lambda,\bm{\Lambda k}')}
a_{+}(\bm{\Lambda k},N)^{\dagger}
a_{-}(\bm{\Lambda k}',N)^{\dagger}
\nonumber\\
&+&
\int d\Gamma(\bm k)  d\Gamma(\bm{k}') ~ 
\psi_{-+}(\bm k,\bm k')
e^{2i\Theta(\Lambda,\bm{\Lambda k})}
e^{-2i\Theta(\Lambda,\bm{\Lambda k}')}
a_{-}(\bm{\Lambda k},N)^{\dagger}
a_{+}(\bm{\Lambda k}',N)^{\dagger}
\nonumber\\
&=&
\int d\Gamma(\bm k)  d\Gamma(\bm{k}') ~ 
\psi_{+-}(\bm{\Lambda k},\bm{ \Lambda k'})
a_{+}(\bm{\Lambda k},N)^{\dagger}
a_{-}(\bm{\Lambda k}',N)^{\dagger}
\nonumber\\
&+&
\int d\Gamma(\bm k)  d\Gamma(\bm{k}') ~ 
\psi_{-+}(\bm{\Lambda k},\bm{ \Lambda k'})
a_{-}(\bm{\Lambda k},N)^{\dagger}
a_{+}(\bm{\Lambda k}',N)^{\dagger}
~=~
\Psi_{11}(N).
\end{eqnarray}
This implies the following transformation rules for the fields:
\begin{eqnarray}
\psi_{+-}(\bm{\Lambda^{-1} k},\bm{ \Lambda^{-1} k'})
&=&
\psi_{+-}(\bm k,\bm k')\label{Psi11+-trans}
e^{2i\Theta(\Lambda,\bm{k})}
e^{-2i\Theta(\Lambda,\bm{k}')}
,
\\
\psi_{-+}(\bm{\Lambda^{-1} k},\bm{ \Lambda^{-1} k'})
&=& 
\psi_{-+}(\bm k,\bm k')\label{Psi11-+trans1}
e^{-2i\Theta(\Lambda,\bm{k})}
e^{2i\Theta(\Lambda,\bm{k}')}
.
\end{eqnarray}
On the other hand from the condition for maximally anti-correlated states in linear basis (\ref{bellcond11}) we get
\begin{eqnarray}\label{Psi11fortheta1}
\psi_{-+}(\bm{\Lambda^{-1} k},\bm{\Lambda^{-1} k}')
&=&
- e^{2i(\theta_{11}(\bm{\Lambda^{-1} k})
-\theta_{11}(\bm{\Lambda^{-1} k}'))}
\psi_{+-}(\bm{\Lambda^{-1} k},\bm{\Lambda^{-1} k}')
\nonumber\\
&=&
- e^{2i(\theta_{11}(\bm{\Lambda^{-1} k})
-\theta_{11}(\bm{\Lambda^{-1} k}'))}
\psi_{+-}(\bm k,\bm k')
e^{2i\Theta(\Lambda,\bm{k})}
e^{-2i\Theta(\Lambda,\bm{k}')}
\nonumber\\
&=&
e^{2i(\theta_{11}(\bm{\Lambda^{-1} k})
-\theta_{11}(\bm{\Lambda^{-1} k}'))}
\psi_{-+}(\bm k,\bm k')
e^{-2i(\theta_{11}(\bm{k})
-\theta_{11}(\bm{k}'))}
e^{2i\Theta(\Lambda,\bm{k})}
e^{-2i\Theta(\Lambda,\bm{k}')}
.
\nonumber\\
\end{eqnarray}
Comparison of (\ref{Psi11-+trans1}) and (\ref{Psi11fortheta1}) implies the transformation rule for the polarization angle under Lorentz transformation:
\begin{eqnarray}\label{transtheta11}
\theta_{11}(\bm{\Lambda^{-1} k})
&=&
\theta_{11}(\bm{k})
-
2\Theta(\Lambda,\bm{k})
.
\end{eqnarray}
We can interpret this as if the polarization angle due to Lorentz transformation is shifted by the Wigner phase. With this condition it can be shown that indeed the field operator $\Psi_{11}(N)$ transforms as a scalar in both bases: linear and circular.
Let us remind ourselves that the Wigner phase depends only on the direction of momentum, not on the frequency, so that all parallel wave vectors correspond to the same rotational angle. This was shown by Caban and Rembieli\'{n}ski in \cite{PCJR03}.
Now taking into account conditions (\ref{bellcond11}) and transformation rules for the null tetrad shown in appendix (\ref{transrule1})-(\ref{transrule1end}) we can write an example of the fields in terms of the null tetrad:
\begin{eqnarray}\label{psi11null}
\psi_{+-}(\bm k,\bm k')
&=&
m_a(\bm k)\bar{m}^a(\bm k'),
\\
\psi_{-+}(\bm k,\bm k')
&=& 
\bar{m}_a(\bm k)m^a(\bm k').
\end{eqnarray}
We can also write the field operator $\Psi_{11}(N)$ with respect to the null tetrad, i.e.
\begin{eqnarray}
\Psi_{11}(N)
&=&
\int d\Gamma(\bm k)  d\Gamma(\bm{k}') ~ 
\left(
m_a(\bm k)\bar{m}^a(\bm k')
a_{+}(\bm k,N)^{\dagger}
a_{-}(\bm k',N)^{\dagger}
+
\bar{m}_a(\bm k)m^a(\bm k')
a_{-}(\bm k,N)^{\dagger}
a_{+}(\bm k',N)^{\dagger}
\right)
\nonumber\\
&=&
2i
\int d\Gamma(\bm k)  d\Gamma(\bm{k}') ~ 
e^{-i (\theta_{11}(\bm k)-\theta_{11}(\bm k'))}
\bar{m}_a(\bm k)m^a(\bm k')
a_{\theta}(\bm k,N)^{\dagger}
a_{\theta'}(\bm k',N)^{\dagger}.
\end{eqnarray}
Under Lorentz transformation we have the following transformation rules for creation operators in linear basis.
\begin{eqnarray}
U(\Lambda,N)\label{Loratheta}
a_\theta(\bm k,N)^{\dagger}
a_{\theta'}(\bm k',N)^{\dagger}
U(\Lambda,N)^{\dagger}&=&
a_\theta(\bm{\Lambda k},N)^{\dagger}
a_{\theta'}(\bm{\Lambda k}',N)^{\dagger}
,
\end{eqnarray}
\begin{eqnarray}\label{Loratheta2}
U(\Lambda,N)
a_\theta(\bm k,N)^{\dagger}
a_{\theta}(\bm k',N)^{\dagger}
U(\Lambda,N)^{\dagger}
&=&
a_\theta(\bm{\Lambda k},N)^{\dagger}
a_{\theta}(\bm{\Lambda k}',N)^{\dagger}
,
\end{eqnarray}
\begin{eqnarray}\label{Loratheta3}
U(\Lambda,N)
a_{\theta '}(\bm k,N)^{\dagger}
a_{\theta '}(\bm k',N)^{\dagger}
U(\Lambda,N)^{\dagger}
&=&
a_{\theta '}(\bm{\Lambda k},N)^{\dagger}
a_{\theta '}(\bm{\Lambda k}',N)^{\dagger}
.
\end{eqnarray}
It should be stressed that, for this calculus, the transformation rule for the polarization angle was taken into account. 
Now let us see how the field operator $\Psi_{11}(N)$ in linear basis transforms under Lorentz transformation. Using the transformation formula (\ref{Loratheta}) we find that
\begin{eqnarray}\label{tildaPsi11}
&&
U(\Lambda,N)
\Psi_{11}(N)
U(\Lambda,N)^{\dagger}
\nonumber\\
&=&
2i
\int d\Gamma(\bm k)  d\Gamma(\bm{k}') ~ 
e^{-i (\theta_{11}(\bm k)-\theta_{11}(\bm k'))}
\bar{m}_a(\bm k)m^a(\bm k')
a_{\theta}(\bm{\Lambda k},N)^{\dagger}
a_{\theta'}(\bm{\Lambda k}',N)^{\dagger}
\nonumber\\
&=&
2i
\int d\Gamma(\bm k)  d\Gamma(\bm{k}') ~ 
e^{-i (\theta_{11}(\bm {\Lambda^{-1}k})-\theta_{11}(\bm {\Lambda^{-1}k}'))}
\bar{m}_a(\bm{\Lambda^{-1}k})m^a(\bm{\Lambda^{-1}k}')
a_{\theta}(\bm{k},N)^{\dagger}
a_{\theta'}(\bm{ k}',N)^{\dagger}
\nonumber\\
&=&
2i
\int d\Gamma(\bm k)  d\Gamma(\bm{k}') ~ 
e^{-i (\theta_{11}(\bm {k})-\theta_{11}(\bm {k}'))}
\bar{m}_a(\bm{k})m^a(\bm{k}')
a_{\theta}(\bm{k},N)^{\dagger}
a_{\theta'}(\bm{k}',N)^{\dagger}.
\end{eqnarray}
As one can see such a state is still maximally correlated after Lorentz transformation. 
This was discussed earlier by M. Czachor in \cite{MC10} and by H. Terashima and M. Ueda in \cite{HTMU0211}. 
The correlation in both bases depends on the relation between the polarization angle $\theta(\bm{k})$ and the fields. 
Conclusion is that to maintain under Lorentz transformations maximal entanglement in EPR-type experiments in both bases, one has to employ momentum dependent polarization functions $\theta(\bm k)$ that compensate the Wigner phase $2\Theta(\Lambda,\bm{k})$.
Now let us consider the field operator $\Psi_{12}(N)$  (\ref{Psi12}). In circular basis we want this field operator to transform under Lorentz transformation as a scalar:
\begin{eqnarray}
&&
U(\Lambda,N)
\Psi_{12}(N)
U(\Lambda,N)^{\dagger}
\nonumber\\
&=&
U(\Lambda,N)
\sum_{s=\pm}
\int d\Gamma(\bm k)  d\Gamma(\bm{k}') ~ 
\psi_{s-s}(\bm k,\bm k')
a_{s}(\bm k,N)^{\dagger}
a_{-s}(\bm k',N)^{\dagger}
U(\Lambda,N)^{\dagger}
\nonumber\\
&=&
\int d\Gamma(\bm k)  d\Gamma(\bm{k}') ~ 
\psi_{+-}(\bm k,\bm k')
e^{-2i\Theta(\Lambda,\bm{\Lambda k})}
e^{2i\Theta(\Lambda,\bm{\Lambda k}')}
a_{+}(\bm{\Lambda k},N)^{\dagger}
a_{-}(\bm{\Lambda k}',N)^{\dagger}
\nonumber\\
&+&
\int d\Gamma(\bm k)  d\Gamma(\bm{k}') ~ 
\psi_{-+}(\bm k,\bm k')
e^{2i\Theta(\Lambda,\bm{\Lambda k})}
e^{-2i\Theta(\Lambda,\bm{\Lambda k}')}
a_{-}(\bm{\Lambda k},N)^{\dagger}
a_{+}(\bm{\Lambda k}',N)^{\dagger}
\nonumber\\
&=&
\int d\Gamma(\bm k)  d\Gamma(\bm{k}') ~ 
\psi_{+-}(\bm{\Lambda k},\bm{ \Lambda k'})
a_{+}(\bm{\Lambda k},N)^{\dagger}
a_{-}(\bm{\Lambda k}',N)^{\dagger}
\nonumber\\
&+&
\int d\Gamma(\bm k)  d\Gamma(\bm{k}') ~ 
\psi_{-+}(\bm{\Lambda k},\bm{ \Lambda k'})
a_{-}(\bm{\Lambda k},N)^{\dagger}
a_{+}(\bm{\Lambda k}',N)^{\dagger}
~=~
\Psi_{12}(N).
\end{eqnarray}
This implies the following transformation rules for the fields:
\begin{eqnarray}
\psi_{+-}(\bm k,\bm k')
e^{2i\Theta(\Lambda,\bm{k})}
e^{-2i\Theta(\Lambda,\bm{k}')}
&=&
\psi_{+-}(\bm{\Lambda^{-1} k},\bm{ \Lambda^{-1} k'}),
\\
\psi_{-+}(\bm k,\bm k')
e^{-2i\Theta(\Lambda,\bm{k})}
e^{2i\Theta(\Lambda,\bm{k}')}
&=& 
\psi_{-+}(\bm{\Lambda^{-1} k},\bm{ \Lambda^{-1} k'}).
\end{eqnarray}
Furthermore, from the condition on the polarization angle and the field (\ref{bellcond12}) we get
\begin{eqnarray}
\psi_{-+}(\bm{\Lambda^{-1} k},\bm{\Lambda^{-1} k}')
&=&
e^{2i(\theta_{12}(\bm{\Lambda^{-1} k})
-\theta_{12}(\bm{\Lambda^{-1} k}'))}
\psi_{+-}(\bm{\Lambda^{-1} k},\bm{\Lambda^{-1} k}')
\nonumber\\
&=&
e^{2i(\theta_{12}(\bm{\Lambda^{-1} k})
-\theta_{12}(\bm{\Lambda^{-1} k}'))}
\psi_{+-}(\bm k,\bm k')
e^{2i\Theta(\Lambda,\bm{k})}
e^{-2i\Theta(\Lambda,\bm{k}')}
\nonumber\\
&=&
e^{2i(\theta_{12}(\bm{\Lambda^{-1} k})
-\theta_{12}(\bm{\Lambda^{-1} k}'))}
\psi_{-+}(\bm k,\bm k')
e^{-2i(\theta_{12}(\bm{k})
-\theta_{12}(\bm{k}'))}
e^{2i\Theta(\Lambda,\bm{k})}
e^{-2i\Theta(\Lambda,\bm{k}')}
,
\nonumber\\
\end{eqnarray}
and this implies the transformation rule for the polarization angle $\theta_{12}(\bm k)$:
\begin{eqnarray}
\theta_{12}(\bm{\Lambda^{-1} k})\label{transruletheta12}
&=&
\theta_{12}(\bm{k})
-
2\Theta(\Lambda,\bm{k}).
\end{eqnarray}
Now we can write the field operator $\Psi_{12}(N)$ (\ref{Psi12}) with respect to the null tetrad
\begin{eqnarray}
\Psi_{12}(N)
&=&
\int d\Gamma(\bm k)  d\Gamma(\bm{k}') ~ 
e^{i (\theta_{12}(\bm k)-\theta_{12}(\bm k'))}
m_a(\bm k)\bar{m}^a(\bm k')
\left(
a_{\theta}(\bm k,N)^{\dagger}
a_{\theta}(\bm k',N)^{\dagger}
+
a_{\theta'}(\bm k,N)^{\dagger}
a_{\theta'}(\bm k',N)^{\dagger}
\right).
\nonumber\\
\end{eqnarray}
Such field holds the scalar field condition also in the linear polarization basis.
\subsection{Transformation properties of states maximally correlated in circular basis}\label{sec:transtriplet}
For the field operator $\Psi_{21}(N)$ (\ref{Psi21}) we first assume the following transformation rule in circular basis
\begin{eqnarray}
&&
U(\Lambda,N)
\Psi_{21}(N)
U(\Lambda,N)^{\dagger}
\nonumber\\
&=&
U(\Lambda,N)
\sum_{s=\mp}
\int d\Gamma(\bm k)  d\Gamma(\bm{k}') ~ 
\psi_{ss}(\bm k,\bm k')
a_{s}(\bm k,N)^{\dagger}
a_{s}(\bm k',N)^{\dagger}
U(\Lambda,N)^{\dagger}
\nonumber\\
&=&
\int d\Gamma(\bm k)  d\Gamma(\bm{k}') ~ 
\psi_{++}(\bm k,\bm k')
e^{2i\Theta(\Lambda,\bm{\Lambda k})}
e^{2i\Theta(\Lambda,\bm{\Lambda k}')}
a_{+}(\bm{\Lambda k},N)^{\dagger}
a_{+}(\bm{\Lambda k}',N)^{\dagger}
\nonumber\\
&+&
\int d\Gamma(\bm k)  d\Gamma(\bm{k}') ~ 
\psi_{--}(\bm k,\bm k')
e^{-2i\Theta(\Lambda,\bm{\Lambda k})}
e^{-2i\Theta(\Lambda,\bm{\Lambda k}')}
a_{-}(\bm{\Lambda k},N)^{\dagger}
a_{-}(\bm{\Lambda k}',N)^{\dagger}
\nonumber\\
&=&
\int d\Gamma(\bm k)  d\Gamma(\bm{k}') ~ 
\psi_{++}(\bm{\Lambda k},\bm{ \Lambda k'})
a_{+}(\bm{\Lambda k},N)^{\dagger}
a_{+}(\bm{\Lambda k}',N)^{\dagger}
\nonumber\\
&+&
\int d\Gamma(\bm k)  d\Gamma(\bm{k}') ~ 
\psi_{--}(\bm{\Lambda k},\bm{ \Lambda k'})
a_{-}(\bm{\Lambda k},N)^{\dagger}
a_{-}(\bm{\Lambda k}',N)^{\dagger}
~=~
\Psi_{21}(N).
\end{eqnarray}
This implies that the fields have to transform as:
\begin{eqnarray}
\psi_{++}(\bm k,\bm k')
e^{2i\Theta(\Lambda,\bm{k})}
e^{2i\Theta(\Lambda,\bm{k}')}
&=&
\psi_{++}(\bm{\Lambda^{-1} k},\bm{ \Lambda^{-1} k'}),
\\
\psi_{--}(\bm k,\bm k')
e^{-2i\Theta(\Lambda,\bm{k})}
e^{-2i\Theta(\Lambda,\bm{k}')}
&=& 
\psi_{--}(\bm{\Lambda^{-1} k},\bm{ \Lambda^{-1} k'}).
\end{eqnarray}
From (\ref{bellcond21}) we get
\begin{eqnarray}
\psi_{--}(\bm{\Lambda^{-1} k},\bm{\Lambda^{-1} k}')
&=&
-
e^{2i(\theta_{21}(\bm{\Lambda^{-1} k})
+\theta_{21}(\bm{\Lambda^{-1} k}'))}
\psi_{++}(\bm{\Lambda^{-1} k},\bm{\Lambda^{-1} k}')
\nonumber\\
&=&
-
e^{2i(\theta_{21}(\bm{\Lambda^{-1} k})
-\theta_{21}(\bm{\Lambda^{-1} k}'))}
\psi_{++}(\bm k,\bm k')
e^{2i\Theta(\Lambda,\bm{k})}
e^{2i\Theta(\Lambda,\bm{k}')}
\nonumber\\
&=&
e^{2i(\theta_{21}(\bm{\Lambda^{-1} k})
+\theta_{21}(\bm{\Lambda^{-1} k}'))}
\psi_{--}(\bm k,\bm k')
e^{-2i(\theta_{21}(\bm{k})
+\theta_{21}(\bm{k}'))}
e^{2i\Theta(\Lambda,\bm{k})}
e^{2i\Theta(\Lambda,\bm{k}')}
.
\nonumber\\
\end{eqnarray}
This implies the following transformation rule for the polarization angle $\theta_{21}(\bm k)$:
\begin{eqnarray}
\theta_{21}(\bm{\Lambda^{-1} k})
&=&
\theta_{21}(\bm{k})\label{transruletheta21}
-
2\Theta(\Lambda,\bm{k}).
\end{eqnarray}
Taking into account conditions (\ref{bellcond21}) and transformation rules for the null tetrad (\ref{transrule1}) - (\ref{transrule1end}), we can write an example of fields in terms of the null tetrad:
\begin{eqnarray}\label{psi21null}
\psi_{++}(\bm k,\bm k')
&=&
m_a(\bm k)m^a(\bm k'),
\\
\psi_{--}(\bm k,\bm k')
&=& 
\bar{m}_a(\bm k)\bar{m}^a(\bm k').
\end{eqnarray}
Now we can write the field operator $\Psi_{21}(N)$ (\ref{Psi21}) with respect to the null tetrad in circular and linear bases
\begin{eqnarray}
\Psi_{21}(N)
&=&
\int d\Gamma(\bm k)  d\Gamma(\bm{k}') ~ 
\left(
m_a(\bm k)m^a(\bm k')
a_{+}(\bm k,N)^{\dagger}
a_{+}(\bm k',N)^{\dagger}
+
\bar{m}_a(\bm k)\bar{m}^a(\bm k')
a_{-}(\bm k,N)^{\dagger}
a_{-}(\bm k',N)^{\dagger}
\right)
\nonumber\\
&=&
-2i
\int d\Gamma(\bm k)  d\Gamma(\bm{k}') ~ 
e^{-i (\theta_{21}(\bm k)+\theta_{21}(\bm k'))}
\bar{m}_a(\bm k)\bar{m}^a(\bm k')
a_{\theta}(\bm k,N)^{\dagger}
a_{\theta'}(\bm k',N)^{\dagger}.
\end{eqnarray}
~~\\
Finally we will consider the  field operator corresponding to the field operator $\Psi_{22}(N)$ (\ref{Psi22}). First we would want this operator in circular basis to transform as a scalar field:
\begin{eqnarray}
&&
U(\Lambda,N)
\Psi_{22}(N)
U(\Lambda,N)^{\dagger}
\nonumber\\
&=&
U(\Lambda,N)
\sum_{s=\pm}
\int d\Gamma(\bm k)  d\Gamma(\bm{k}') ~ 
\psi_{ss}(\bm k,\bm k')
a_{s}(\bm k,N)^{\dagger}
a_{s}(\bm k',N)^{\dagger}
U(\Lambda,N)^{\dagger}
\nonumber\\
&=&
\int d\Gamma(\bm k)  d\Gamma(\bm{k}') ~ 
\psi_{++}(\bm k,\bm k')
e^{2i\Theta(\Lambda,\bm{\Lambda k})}
e^{2i\Theta(\Lambda,\bm{\Lambda k}')}
a_{+}(\bm{\Lambda k},N)^{\dagger}
a_{+}(\bm{\Lambda k}',N)^{\dagger}
\nonumber\\
&+&
\int d\Gamma(\bm k)  d\Gamma(\bm{k}') ~ 
\psi_{--}(\bm k,\bm k')
e^{-2i\Theta(\Lambda,\bm{\Lambda k})}
e^{-2i\Theta(\Lambda,\bm{\Lambda k}')}
a_{-}(\bm{\Lambda k},N)^{\dagger}
a_{-}(\bm{\Lambda k}',N)^{\dagger}
\nonumber\\
&=&
\int d\Gamma(\bm k)  d\Gamma(\bm{k}') ~ 
\psi_{++}(\bm{\Lambda k},\bm{ \Lambda k'})
a_{+}(\bm{\Lambda k},N)^{\dagger}
a_{+}(\bm{\Lambda k}',N)^{\dagger}
\nonumber\\
&+&
\int d\Gamma(\bm k)  d\Gamma(\bm{k}') ~ 
\psi_{--}(\bm{\Lambda k},\bm{ \Lambda k'})
a_{-}(\bm{\Lambda k},N)^{\dagger}
a_{-}(\bm{\Lambda k}',N)^{\dagger}
~=~
\Psi_{22}(N).
\end{eqnarray}
This implies the following transformation rule on the field
\begin{eqnarray}
\psi_{++}(\bm k,\bm k')
e^{2i\Theta(\Lambda,\bm{k})}
e^{2i\Theta(\Lambda,\bm{k}')}
&=&
\psi_{++}(\bm{\Lambda^{-1} k},\bm{ \Lambda^{-1} k'}),
\nonumber\\
\psi_{--}(\bm k,\bm k')
e^{-2i\Theta(\Lambda,\bm{k})}
e^{-2i\Theta(\Lambda,\bm{k}')}
&=& 
\psi_{--}(\bm{\Lambda^{-1} k},\bm{ \Lambda^{-1} k'}).
\end{eqnarray}
From (\ref{bellcond22}) we get
\begin{eqnarray}
\psi_{--}(\bm{\Lambda^{-1} k},\bm{\Lambda^{-1} k}')
&=&
e^{2i(\theta_{22}(\bm{\Lambda^{-1} k})
+\theta_{22}(\bm{\Lambda^{-1} k}'))}
\psi_{++}(\bm{\Lambda^{-1} k},\bm{\Lambda^{-1} k}')
\nonumber\\
&=&
e^{2i(\theta_{22}(\bm{\Lambda^{-1} k})
-\theta_{22}(\bm{\Lambda^{-1} k}'))}
\psi_{++}(\bm k,\bm k')
e^{2i\Theta(\Lambda,\bm{k})}
e^{2i\Theta(\Lambda,\bm{k}')}
\nonumber\\
&=&
e^{2i(\theta_{22}(\bm{\Lambda^{-1} k})
+\theta_{22}(\bm{\Lambda^{-1} k}'))}
\psi_{--}(\bm k,\bm k')
e^{-2i(\theta_{22}(\bm{k})
+\theta_{22}(\bm{k}'))}
e^{2i\Theta(\Lambda,\bm{k})}
e^{2i\Theta(\Lambda,\bm{k}')}
.
\nonumber\\
\end{eqnarray}
This implies the following transformation rule for the polarization angle $\theta_{22}(\bm k)$:
\begin{eqnarray}
\theta_{22}(\bm{\Lambda^{-1} k})
&=&
\theta_{22}(\bm{k})\label{transruletheta22}
-
2\Theta(\Lambda,\bm{k}).
\end{eqnarray}
Now we can write the  field operator $\Psi_{22}(N)$ (\ref{Psi22}) with respect to the null tetrad
\begin{eqnarray}
\Psi_{22}(N)
&=&
\int d\Gamma(\bm k)  d\Gamma(\bm{k}') ~ 
\left(
m_a(\bm k)m^a(\bm k')
a_{+}(\bm k,N)^{\dagger}
a_{+}(\bm k',N)^{\dagger}
+
\bar{m}_a(\bm k)\bar{m}^a(\bm k')
a_{-}(\bm k,N)^{\dagger}
a_{-}(\bm k',N)^{\dagger}
\right)
\nonumber\\
&=&
\int d\Gamma(\bm k)  d\Gamma(\bm{k}') ~ 
e^{i (\theta_{22}(\bm k)+\theta_{22}(\bm k'))}
m_a(\bm k)m^a(\bm k')
\left(
a_{\theta}(\bm k,N)^{\dagger}
a_{\theta}(\bm k',N)^{\dagger}
-
a_{\theta'}(\bm k,N)^{\dagger}
a_{\theta'}(\bm k',N)^{\dagger}
\right)
\nonumber\\
&=&
\int d\Gamma(\bm k)  d\Gamma(\bm{k}') ~ 
e^{-i (\theta_{22}(\bm k)+\theta_{22}(\bm k'))}
\bar{m}_a(\bm k)\bar{m}^a(\bm k')
\left(
a_{\theta}(\bm k,N)^{\dagger}
a_{\theta}(\bm k',N)^{\dagger}
-
a_{\theta'}(\bm k,N)^{\dagger}
a_{\theta'}(\bm k',N)^{\dagger}
\right).
\nonumber\\
\end{eqnarray}
Under Lorentz transformation field operators $\Psi_{21}(N)$ and $\Psi_{22}(N)$ holds the scalar field condition in both polarization bases. 
Without the transformation rule on the polarization angle such states no longer would have maintained the maximal correlation. \\
\section{Observables in EPR-type experiments}\label{sec:detection}
\subsection{Yes-no observable}\label{sec:detectionEPR}
Let us first define a yes-no observable for the 
linear polarizations:
\begin{eqnarray}\label{Y-N}
Y_{\alpha}(\bm l,N)
=
n_{\alpha}(\bm l,N)
-
n_{\alpha'}(\bm l,N).
\end{eqnarray}
Here $n_{\alpha}(\bm l,N)$ is the number operator for $\alpha$ oriented polarizations in $N$-oscillator reducible representation. In (\ref{Y-N}) we use the definition ($\ref{nIIIN}$) for the number operator. 
This observable may describe measurement on detectors oriented in $\alpha$ direction in EPR-type experiments, and $\alpha'$ is denoted here as $\alpha'=\alpha+\frac{\pi}{2}$. 
We can write the yes-no observable also with respect to circular polarizations, i.e.
\begin{eqnarray}\label{YNcirc}
Y_{\alpha}(\bm l,N)
&=&
\sum_{n=1}^{N}
\sum_{s=\pm}
\left(
e^{2is\alpha}
|\bm l\rangle\langle\bm l|\otimes
a_{-s}^{\dagger}a_{s}
\right)_.^{(n)}
\end{eqnarray}
Let us note that an observable so defined measures the polarization angle with respect to the circular polarization basis (left-handed or right-handed). This can be seen when in (\ref{YNcirc}) we change the sing of the $\alpha$ angle, i.e.
\begin{eqnarray}\label{YNcirc-alpha}
Y_{-\alpha}(\bm l,N)
&=&
\sum_{n=1}^{N}
\sum_{s=\pm}
\left(
e^{2is\alpha}
|\bm l\rangle\langle\bm l|\otimes
a_{s}^{\dagger}a_{-s}
\right)_.^{(n)}
\end{eqnarray}
 This remark is important for further interpretation of the correlation functions, where in next sections for anti-correlated in circular basis states we get a $\cos(\beta-\alpha)$ term and for correlated in circular basis states a $\cos(\beta+\alpha)$ term.
For a more realistic case the localization of the photon 
detector leads to a momentum solid angle spread 
$\bm l\in \Omega$, and this is why we will consider the following observable 
\begin{eqnarray}\label{YalphaN}
Y_{\alpha}(N)
=
\int\limits_{\Omega} d\Gamma(\bm l)~
Y_{\alpha}(\bm l,N).
\end{eqnarray}
It should be stressed here that the $\alpha$ angle in detectors is fixed for all momentum values of the photon field. This is  important for the relativistic background, because it is assumed here that the polarization angle for linearly polarized fields depends on the momentum and is shifted due to the Wigner phase under Lorentz transformation. Let us remind ourselves that this dependence on momentum is necessary for maintaining Lorentz invariance for all of the four Bell states. On the other hand we know that the Wigner phase depends only on the direction of the momentum, so for parallel wave vectors this would not effect the detection.\\
\\
Observable (\ref{YalphaN}) doesn't always give eigenvalue +1 for one photon fields polarized under $\alpha$ angle and $-$1 for fields polarized under angle $\alpha'$. Taking under consideration this eigenvalue problem we see that
\begin{eqnarray}\label{}
Y_{\alpha}(N)
|\Psi_{\alpha}(N,1) \rangle
&=&
\int\limits_{\Omega} d\Gamma(\bm l)~
Y_{\alpha}(\bm l,N)
\int d \Gamma(\bm k)~
\Psi(\bm{k},n_{\alpha})
a_{\alpha}^{\dag}(\bm k,N)
|O(N)\rangle
\nonumber\\
&=&
\frac{1}{\sqrt{2}}
\int\limits_{\Omega} d\Gamma(\bm l)~
\Psi(\bm{l},n_{\alpha})
\sum_{n=1}^{N}
\left(
\sum_{s=\pm}
|\bm l\rangle\langle\bm l|
\otimes
a_{s}^{\dagger}
e^{is\alpha(\bm l)}
e^{-2is\alpha}
\right)^{(n)}
|O(N)\rangle
.
\end{eqnarray}
So first we have to assume that the support of the wave function $\Psi(\bm l,n_\alpha)$ is embedded in $\Omega$ and further within this angle the dependence on momentum of the polarization can be neglected, so  $\alpha(\bm l)=\alpha$. Then the eigenvalue will be +1, i.e. 
\begin{eqnarray}\label{}
Y_{\alpha}(N)
|\Psi_{\alpha}(N,1) \rangle
&=&
+1
|\Psi_{\alpha}(N,1)\rangle.
\end{eqnarray}
For fields polarized under a perpendicular angle, in analogy to the previous calculus, we get
\begin{eqnarray}\label{}
Y_{\alpha}(N)
|\Psi_{\alpha'}(N,1) \rangle
&=&
\int\limits_{\Omega} d\Gamma(\bm l)~
Y_{\alpha}(\bm l,N)
\int d \Gamma(\bm k)~
\Psi(\bm{k},n_{\alpha'})
a_{\alpha'}^{\dag}(\bm k,N)
|O(N)\rangle
\nonumber\\
&=&
-
\frac{1}{\sqrt{2}}
\int\limits_{\Omega} d\Gamma(\bm l)~
\Psi(\bm{l},n_{\alpha'})
\sum_{n=1}^{N}
\left(
\sum_{s=\pm}
|\bm l\rangle\langle\bm l|
\otimes
a_{s}^{\dagger}
e^{is\alpha'(\bm l)}
e^{-2is\alpha'}
\right)^{(n)}
|O(N)\rangle
.
\end{eqnarray}
Again, assuming the wave function is all concentrated within $\Omega$ and within this angle spread the dependence on momentum of the polarization angle can be neglected, we get
\begin{eqnarray}\label{}
Y_{\alpha}(N)
|\Psi_{\alpha'}(N,1) \rangle
&=&
-1
|\Psi_{\alpha'}(N,1)\rangle.
\end{eqnarray}
\subsection{Correlation function for two-photon states}\label{sec:detectiontwophoton}
Now let us consider two observers in the same inertial frame. Alice measures $\alpha$ oriented photons and Bob $\beta$ oriented ones. Their observables are $Y_{\alpha}(N)$ and $Y_{\beta}(N)$ respectively.
In a more realistic case the localization of the photon 
detectors leads to momentum solid angle spreads 
$\bm l\in \Omega, ~\bm l'\in \Omega'$ respectively. 
Then the normalized correlation function for an arbitrary two state photon will be then given by:
\begin{eqnarray}
\frac{
\langle O(N)|
\Psi(N)^{\dagger}
Y_\beta (N)
Y_\alpha (N)
\Psi(N)
|O(N)\rangle
}
{
\langle O(N)|
\Psi(N)^{\dagger}
\Psi(N)
|O(N)\rangle}
.
\end{eqnarray}
First we will consider the two-photon field operator $\Psi(N)$ (\ref{PsiNs}). The unnormalized ERP average can be written in the form
\begin{eqnarray}\label{readyscalarPsiYYPsi}
&&
\langle O(N)|
\Psi(N)^{\dagger}
Y_\beta (N)
Y_\alpha (N)
\Psi(N)
|O(N)\rangle
\nonumber\\
&=&
4
\sum_{ss'=\pm}
\int\limits_{\Omega} d\Gamma(\bm{l})~ 
\int\limits_{\Omega'} d\Gamma(\bm{l}')~ 
e^{-2i(s\beta+s'\alpha)}
\bar{\psi}_{ss'}(\bm l,\bm l')
\psi_{-s'-s}(\bm l',\bm l)
\langle O(N)|
I(\bm l,N)I(\bm l',N)
|O(N)\rangle
\nonumber\\
&+&
4
\sum_{ss'=\pm}
\int\limits_{\Omega} d\Gamma(\bm{l})~ 
\int\limits_{\Omega'} d\Gamma(\bm{l}')~ 
\int d\Gamma(\bm k)  ~ 
e^{-2i(s\beta-s\alpha)}
\bar{\psi}_{ss'}(\bm l,\bm k)
\psi_{ss'}(\bm l',\bm k)
\delta_{\Gamma}(\bm l, \bm l')
\langle O(N)|
I(\bm k,N)I(\bm l,N)
|O(N)\rangle.
\nonumber\\
\end{eqnarray}
In the case of disjoint detectors, i.e. $\Omega \cap \Omega' = \emptyset$, just one part of (\ref{readyscalarPsiYYPsi}) has contribution to the EPR average, so that
\begin{eqnarray}
&&
\langle O(N)|
\Psi(N)^{\dagger}
Y_\beta (N)
Y_\alpha (N)
\Psi(N)
|O(N)\rangle
\nonumber\\
&=&
\frac{4(N-1)}{N}
\sum_{ss'=\pm}
\int\limits_{\Omega} d\Gamma(\bm{l})~ 
\int\limits_{\Omega'} d\Gamma(\bm{l}')~ 
e^{-2i(s\beta+s'\alpha)}
\bar{\psi}_{ss'}(\bm l,\bm l')
\psi_{-s'-s}(\bm l',\bm l)
Z(\bm l)Z(\bm l').
\end{eqnarray}
Now let us use the symmetry condition (\ref{symcond}) and take a closer look at part:
\begin{eqnarray}\label{eprsym}
&&
\sum_{ss'=\pm}
e^{-2i(s\beta+s'\alpha)}
\bar{\psi}_{ss'}(\bm l,\bm l')
\psi_{-s-s'}(\bm l,\bm l')
~=~
\sum_{ss'=\pm}
e^{2i(s\beta+s'\alpha)}
\bar{\psi}_{-s-s'}(\bm l,\bm l')
\psi_{ss'}(\bm l,\bm l')
\nonumber\\
&=&
2 \cos 2(\beta+\alpha)
\Re\left(
\bar{\psi}_{++}(\bm l,\bm l')
\psi_{--}(\bm l,\bm l')
\right)
+
2\sin 2(\beta+\alpha)
\Im\left(
\bar{\psi}_{++}(\bm l,\bm l')
\psi_{--}(\bm l,\bm l')
\right)
\nonumber\\
&+&
2\cos 2(\beta-\alpha)
\Re\left(
\bar{\psi}_{+-}(\bm l,\bm l')
\psi_{-+}(\bm l,\bm l')
\right)
+
2\sin 2(\beta-\alpha)
\Im\left(
\bar{\psi}_{+-}(\bm l,\bm l')
\psi_{-+}(\bm l,\bm l')
\right).
\end{eqnarray}
This part is purely real and after some basic manipulations we come to the form (\ref{eprsym}). This calculus was done mostly to bring up the $\cos 2(\beta\pm\alpha)$ part, known from literature. 
So the unnormalized EPR average for an arbitrary two-photon state can be written in the form
\begin{eqnarray}
&&
\langle O(N)|
\Psi(N)^{\dagger}
Y_\beta (N)
Y_\alpha (N)
\Psi(N)
|O(N)\rangle
\nonumber\\
&=&
\frac{8(N-1)}{N}
\cos 2(\beta+\alpha)
\int\limits_{\Omega} d\Gamma(\bm{l})~ 
\int\limits_{\Omega'} d\Gamma(\bm{l}')~ 
\Re\left(
\bar{\psi}_{++}(\bm l,\bm l')
\psi_{--}(\bm l,\bm l')
\right)
Z(\bm l)Z(\bm l')
\nonumber\\
&+&
\frac{8(N-1)}{N}
\sin 2(\beta+\alpha)
\int\limits_{\Omega} d\Gamma(\bm{l})~ 
\int\limits_{\Omega'} d\Gamma(\bm{l}')~ 
\Im\left(
\bar{\psi}_{++}(\bm l,\bm l')
\psi_{--}(\bm l,\bm l')
\right)
Z(\bm l)Z(\bm l')
\nonumber\\
&+&
\frac{8(N-1)}{N}
\cos 2(\beta-\alpha)
\int\limits_{\Omega} d\Gamma(\bm{l})~ 
\int\limits_{\Omega'} d\Gamma(\bm{l}')~ 
\Re\left(
\bar{\psi}_{+-}(\bm l,\bm l')
\psi_{-+}(\bm l,\bm l')
\right)
Z(\bm l)Z(\bm l')
\nonumber\\
&+&
\frac{8(N-1)}{N}
\sin 2(\beta-\alpha)
\int\limits_{\Omega} d\Gamma(\bm{l})~ 
\int\limits_{\Omega'} d\Gamma(\bm{l}')~ 
\Im\left(
\bar{\psi}_{+-}(\bm l,\bm l')
\psi_{-+}(\bm l,\bm l')
\right)Z(\bm l)Z(\bm l').
\end{eqnarray}
\subsection{Correlation function for maximally anti-correlated in circular polarization basis states}\label{sec:detectionantycorr}
Now let us consider the correlation function for Bell states, starting from the maximally anti-correlated in circular polarization basis field operator $\Psi_1(N)$ (\ref{Psi1}).
In a realistic case, when the localization of the photon 
detectors leads to momentum solid angle spreads 
$\bm l\in \Omega, ~\bm l'\in \Omega'$ and for disjoint detectors 
$\Omega \cap \Omega' = \emptyset$, the correlation function reads:
\begin{eqnarray}
&&
\langle O(N)|
\Psi_{1}(N)^{\dagger}
Y_\beta (N)
Y_\alpha (N)
\Psi_{1}(N)
|O(N)\rangle
\nonumber\\
&=&
\frac{8(N-1)}{N}
\cos 2(\beta-\alpha)
\int\limits_{\Omega} d\Gamma(\bm{l})~ 
\int\limits_{\Omega'} d\Gamma(\bm{l}')~ 
\Re\left(
\bar{\psi}_{+-}(\bm l,\bm l')
\psi_{-+}(\bm l,\bm l')
\right)
Z(\bm l)Z(\bm l')
\nonumber\\
&+&
\frac{8(N-1)}{N}
\sin 2(\beta-\alpha)
\int\limits_{\Omega} d\Gamma(\bm{l})~ 
\int\limits_{\Omega'} d\Gamma(\bm{l}')~ 
\Im\left(
\bar{\psi}_{+-}(\bm l,\bm l')
\psi_{-+}(\bm l,\bm l')
\right)Z(\bm l)Z(\bm l').
\end{eqnarray}
For the field operator $\Psi_{11}(N)$ we will use the  condition on the field and polarization angle (\ref{bellcond11}), and then the EPR average can be written in the form 
\begin{eqnarray}
&&
\langle O(N)|
\Psi_{11}(N)^{\dagger}
Y_\beta (N)
Y_\alpha (N)
\Psi_{11}(N)
|O(N)\rangle
\nonumber\\
&=&
\frac{-8(N-1)}{N}
\int\limits_{\Omega} d\Gamma(\bm{l})~ 
\int\limits_{\Omega'} d\Gamma(\bm{l}')~ 
\cos 2(\beta-\alpha-\theta_{11}(\bm l)+\theta_{11}(\bm l'))
|\psi_{-+}(\bm l,\bm l')|^2
Z(\bm l)Z(\bm l').
\end{eqnarray}
Furthermore, the normalized EPR average reads
\begin{eqnarray}\label{epr11}
&&
\frac{
\langle O(N)|
\Psi_{11}(N)^{\dagger}
Y_\beta (N)
Y_\alpha (N)
\Psi_{11}(N)
|O(N)\rangle
}
{
\langle O(N)|
\Psi_{11}(N)^{\dagger}
\Psi_{11}(N)
|O(N)\rangle}
\nonumber\\
&=&
\displaystyle{
\frac{
-2(N-1)
\int_{\Omega} d\Gamma(\bm{l})~ 
\int_{\Omega'} d\Gamma(\bm{l}')~ 
\cos 2(\beta-\alpha-\theta_{11}(\bm l)+\theta_{11}(\bm l'))
|\psi_{-+}(\bm l,\bm l')|^2
Z(\bm l)Z(\bm l')
}{
\int d\Gamma(\bm k) 
|\psi_{+-}(\bm k,\bm k)|^2
Z(\bm k)
+
(N-1)
\int d\Gamma(\bm k)  d\Gamma(\bm{k}')
|\psi_{+-}(\bm k,\bm k')|^2
Z(\bm k)Z(\bm k')
}
}.
\end{eqnarray}
Now let us use the explicit values of $\psi_{+-}(\bm k, \bm k')$ (\ref{psi11null}) taken from the Lorentz covariance condition of such two-photon field
\begin{eqnarray}\label{epr11}
&&
\frac{
\langle O(N)|
\Psi_{11}(N)^{\dagger}
Y_\beta (N)
Y_\alpha (N)
\Psi_{11}(N)
|O(N)\rangle
}
{
\langle O(N)|
\Psi_{11}(N)^{\dagger}
\Psi_{11}(N)
|O(N)\rangle}
\nonumber\\
&=&
\displaystyle{
\frac{
-2(N-1)
\int_{\Omega} d\Gamma(\bm{l})~ 
\int_{\Omega'} d\Gamma(\bm{l}')~ 
\cos 2(\beta-\alpha-\theta_{11}(\bm l)+\theta_{11}(\bm l'))
|m_a(\bm l)\bar{m}^a(\bm l')|^2
Z(\bm l)Z(\bm l')
}{
1
+
(N-1)
\int d\Gamma(\bm k)  d\Gamma(\bm{k}')
|m_a(\bm k)\bar{m}^a(\bm k')|^2
Z(\bm k)Z(\bm k')
}
}.
\end{eqnarray}
We can see that  the term $
\int d\Gamma(\bm k) 
|\psi_{+-}(\bm k,\bm k)|^2
Z(\bm k)=1$ and this makes the normalized EPR average dependent on the $N$ parameter, while in the $N\to\infty$ limit it takes the form
\begin{eqnarray}\label{epr11term}
&&
\lim_{N\to\infty}
\frac{
\langle O(N)|
\Psi_{11}(N)^{\dagger}
Y_\beta (N)
Y_\alpha (N)
\Psi_{11}(N)
|O(N)\rangle
}
{
\langle O(N)|
\Psi_{11}(N)^{\dagger}
\Psi_{11}(N)
|O(N)\rangle}
\nonumber\\
&=&
\displaystyle{
\frac{
-2
\int_{\Omega} d\Gamma(\bm{l})~ 
\int_{\Omega'} d\Gamma(\bm{l}')~ 
\cos 2(\beta-\alpha-\theta_{11}(\bm l)+\theta_{11}(\bm l'))
|m_a(\bm l)\bar{m}^a(\bm l')|^2
Z(\bm l)Z(\bm l')
}{
\int d\Gamma(\bm k)  d\Gamma(\bm{k}')
|m_a(\bm k)\bar{m}^a(\bm k')|^2
Z(\bm k)Z(\bm k')
}
}.
\end{eqnarray}
For the field operator $\Psi_{12}(N)$  (\ref{Psi12}), we will use the (\ref{bellcond12}) condition on the field and the polarization angle, so that
\begin{eqnarray}
&&
\langle O(N)|
\Psi_{12}(N)^{\dagger}
Y_\beta (N)
Y_\alpha (N)
\Psi_{12}(N)
|O(N)\rangle
\nonumber\\
&=&
\frac{8(N-1)}{N}
\int_{\Omega} d\Gamma(\bm{l})~ 
\int_{\Omega'} d\Gamma(\bm{l}')~ 
\cos 2(\beta-\alpha
-\theta_{12}(\bm l)+\theta_{12}(\bm l'))
|\psi_{-+}(\bm l,\bm l')|^2
Z(\bm l)Z(\bm l').
\end{eqnarray}
Then the normalized EPR average for $\Psi_{12}(N)$ states reads
\begin{eqnarray}\label{epr12}
&&
\frac{
\langle O(N)|
\Psi_{12}(N)^{\dagger}
Y_\beta (N)
Y_\alpha (N)
\Psi_{12}(N)
|O(N)\rangle
}
{
\langle O(N)|
\Psi_{12}(N)^{\dagger}
\Psi_{12}(N)
|O(N)\rangle}
\nonumber\\
&=&
\displaystyle{
\frac{
2(N-1)
\int_{\Omega} d\Gamma(\bm{l})~ 
\int_{\Omega'} d\Gamma(\bm{l}')~ 
\cos 2(\beta-\alpha-\theta_{12}(\bm l)+\theta_{12}(\bm l'))
|\psi_{-+}(\bm l,\bm l')|^2
Z(\bm l)Z(\bm l')
}{
\int d\Gamma(\bm k) 
|\psi_{+-}(\bm k,\bm k)|^2
Z(\bm k)
+
(N-1)
\int d\Gamma(\bm k)  d\Gamma(\bm{k}')
|\psi_{+-}(\bm k,\bm k')|^2
Z(\bm k)Z(\bm k')
}
},
\end{eqnarray}
and similarly for the explicit values of $\psi_{+-}(\bm k, \bm k')$ we get
\begin{eqnarray}\label{epr12}
&&
\frac{
\langle O(N)|
\Psi_{12}(N)^{\dagger}
Y_\beta (N)
Y_\alpha (N)
\Psi_{12}(N)
|O(N)\rangle
}
{
\langle O(N)|
\Psi_{12}(N)^{\dagger}
\Psi_{12}(N)
|O(N)\rangle}
\nonumber\\
&=&
\displaystyle{
\frac{
2(N-1)
\int_{\Omega} d\Gamma(\bm{l})~ 
\int_{\Omega'} d\Gamma(\bm{l}')~ 
\cos 2(\beta-\alpha-\theta_{12}(\bm l)+\theta_{12}(\bm l'))
|m_a(\bm l)\bar{m}^{a}(\bm l')|^2
Z(\bm l)Z(\bm l')
}{
1
+
(N-1)
\int d\Gamma(\bm k)  d\Gamma(\bm{k}')
|m_a(\bm k)\bar{m}^{a}(\bm k')|^2
Z(\bm k)Z(\bm k')
}
}.
\end{eqnarray}
Also the normalized EPR average for $\Psi_{12}(N)$ becomes dependent on the $N$ parameter and in the ${N\to\infty}$ limit becomes
\begin{eqnarray}\label{epr12term}
&&
\lim_{N\to\infty}
\frac{
\langle O(N)|
\Psi_{12}(N)^{\dagger}
Y_\beta (N)
Y_\alpha (N)
\Psi_{12}(N)
|O(N)\rangle
}
{
\langle O(N)|
\Psi_{12}(N)^{\dagger}
\Psi_{12}(N)
|O(N)\rangle}
\nonumber\\
&=&
\displaystyle{
\frac{
2
\int_{\Omega} d\Gamma(\bm{l})~ 
\int_{\Omega'} d\Gamma(\bm{l}')~ 
\cos 2(\beta-\alpha-\theta_{12}(\bm l)+\theta_{12}(\bm l'))
|m_a(\bm l)\bar{m}^{a}(\bm l')|^2
Z(\bm l)Z(\bm l')
}{
\int d\Gamma(\bm k)  d\Gamma(\bm{k}')
|m_a(\bm k)\bar{m}^{a}(\bm k')|^2
Z(\bm k)Z(\bm k')
}
}.
\end{eqnarray}
Furthermore, using relation on the polarization angles $\theta_{11}(\bm k)$ and $\theta_{12}(\bm k)$ (\ref{rel11-12}) we see that
\begin{eqnarray}
\frac{
\langle O(N)|
\Psi_{11}(N)^{\dagger}
Y_\beta (N)
Y_\alpha (N)
\Psi_{11}(N)
|O(N)\rangle
}
{
\langle O(N)|
\Psi_{11}(N)^{\dagger}
\Psi_{11}(N)
|O(N)\rangle}
&=&
\frac{
\langle O(N)|
\Psi_{12}(N)^{\dagger}
Y_\beta (N)
Y_\alpha (N)
\Psi_{12}(N)
|O(N)\rangle
}
{
\langle O(N)|
\Psi_{12}(N)^{\dagger}
\Psi_{12}(N)
|O(N)\rangle}.
\end{eqnarray}
\subsection{Correlation function for maximally correlated in circular polarization basis states}\label{sec:detectioncorr}
Now we will follow the same calculations as from the previous section, only this time for the field operators corresponding to states maximally correlated in circular basis. 
For such field operators let us consider $\Psi_2(N)$ (\ref{Psi2}).
Again we are assuming a realistic case when the localization of the photon 
detectors leads to momentum solid angle spreads 
$\bm l\in \Omega, ~\bm l'\in \Omega'$ and disjoint detectors 
$\Omega \cap \Omega' = \emptyset$. Then the correlation function reads:
\begin{eqnarray}
&&
\langle O(N)|
\Psi_2(N)^{\dagger}
Y_\beta (N)
Y_\alpha (N)
\Psi_2(N)
|O(N)\rangle
\hspace{8.7 cm}
\nonumber\\
&=&
\frac{8(N-1)}{N}
\cos 2(\beta+\alpha)
\int\limits_{\Omega} d\Gamma(\bm{l})~ 
\int\limits_{\Omega'} d\Gamma(\bm{l}')~ 
\Re\left(
\bar{\psi}_{++}(\bm l,\bm l')
\psi_{--}(\bm l,\bm l')
\right)
Z(\bm l)Z(\bm l')
\nonumber\\
&+&
\frac{8(N-1)}{N}
\sin 2(\beta+\alpha)
\int\limits_{\Omega} d\Gamma(\bm{l})~ 
\int\limits_{\Omega'} d\Gamma(\bm{l}')~ 
\Im\left(
\bar{\psi}_{++}(\bm l,\bm l')
\psi_{--}(\bm l,\bm l')
\right)
Z(\bm l)Z(\bm l').
\end{eqnarray}
For the field operator $\Psi_{21}(N)$ (\ref{Psi21}), we will use the (\ref{bellcond21}) condition on the field and polarization angle
\begin{eqnarray}
&&
\langle O(N)|
\Psi_{21}(N)^{\dagger}
Y_\beta (N)
Y_\alpha (N)
\Psi_{21}(N)
|O(N)\rangle
\hspace{8.5 cm}
\nonumber\\
&=&
\frac{-8(N-1)}{N}
\int\limits_{\Omega} d\Gamma(\bm{l})~ 
\int\limits_{\Omega'} d\Gamma(\bm{l}')~ 
\cos 2(\beta+\alpha
-\theta_{21}(\bm l)-\theta_{21}(\bm l'))
|\psi_{--}(\bm l,\bm l')|^2
Z(\bm l)Z(\bm l')
.
\end{eqnarray}
Then the normalized EPR average for $\Psi_{21}(N)$ reads
\begin{eqnarray}\label{epr21}
&&
\frac{
\langle O(N)|
\Psi_{21}(N)^{\dagger}
Y_\beta (N)
Y_\alpha (N)
\Psi_{21}(N)
|O(N)\rangle
}
{
\langle O(N)|
\Psi_{21}(N)^{\dagger}
\Psi_{21}(N)
|O(N)\rangle}
\nonumber\\
&=&
\displaystyle{
\frac{
-2(N-1)
\int_{\Omega} d\Gamma(\bm{l})~ 
\int_{\Omega'} d\Gamma(\bm{l}')~ 
\cos 2(\beta+\alpha-\theta_{21}(\bm l)-\theta_{21}(\bm l'))
|\psi_{--}(\bm l,\bm l')|^2
Z(\bm l)Z(\bm l')
}{
\int d\Gamma(\bm k) 
|\psi_{--}(\bm k,\bm k)|^2
Z(\bm k)
+
(N-1)
\int d\Gamma(\bm k)  d\Gamma(\bm{k}')
|\psi_{--}(\bm k,\bm k')|^2
Z(\bm k)Z(\bm k')
}
}.
\end{eqnarray}
Using the explicit value of  $\psi_{--}(\bm k,\bm k')$ and knowing that $\psi_{--}(\bm k,\bm k)=0$, we come to the EPR average that does not depend on the $N$ parameter, i.e. 
\begin{eqnarray}\label{epr21null}
&&
\frac{
\langle O(N)|
\Psi_{21}(N)^{\dagger}
Y_\beta (N)
Y_\alpha (N)
\Psi_{21}(N)
|O(N)\rangle
}
{
\langle O(N)|
\Psi_{21}(N)^{\dagger}
\Psi_{21}(N)
|O(N)\rangle}
\nonumber\\
&=&
\displaystyle{
\frac{
-2
\int_{\Omega} d\Gamma(\bm{l})~ 
\int_{\Omega'} d\Gamma(\bm{l}')~ 
\cos 2(\beta+\alpha-\theta_{21}(\bm l)-\theta_{21}(\bm l'))
|m_a(\bm l)m^a(\bm l')|^2
Z(\bm l)Z(\bm l')
}{
\int d\Gamma(\bm k)  d\Gamma(\bm{k}')
|m_a(\bm k)m^a(\bm k')|^2
Z(\bm k)Z(\bm k')
}
}.
\end{eqnarray}
For the field operator $\Psi_{22}(N)$ (\ref{Psi22}), we will use the condition on the field and the polarization angle (\ref{bellcond22}), so that the unnormalized EPR average can be written in the form
\begin{eqnarray}
&&
\langle O(N)|
\Psi_{22}(N)^{\dagger}
Y_\beta (N)
Y_\alpha (N)
\Psi_{22}(N)
|O(N)\rangle
\hspace{8.5 cm}
\nonumber\\
&=&
\frac{8(N-1)}{N}
\int\limits_{\Omega} d\Gamma(\bm{l})~ 
\int\limits_{\Omega'} d\Gamma(\bm{l}')~ 
\cos 2(\beta+\alpha-\theta_{22}(\bm l)-\theta_{22}(\bm l'))
|\psi_{--}(\bm l,\bm l')|^2
Z(\bm l)Z(\bm l')
.
\end{eqnarray}
Then the normalized EPR average for $\Psi_{22}(N)$ Bell state reads
\begin{eqnarray}\label{epr22}
&&
\frac{
\langle O(N)|
\Psi_{22}(N)^{\dagger}
Y_\beta (N)
Y_\alpha (N)
\Psi_{22}(N)
|O(N)\rangle
}
{
\langle O(N)|
\Psi_{22}(N)^{\dagger}
\Psi_{22}(N)
|O(N)\rangle}
\nonumber\\
&=&
\displaystyle{
\frac{
2(N-1)
\int_{\Omega} d\Gamma(\bm{l})~ 
\int_{\Omega'} d\Gamma(\bm{l}')~ 
\cos 2(\beta+\alpha-\theta_{22}(\bm l)-\theta_{22}(\bm l'))
|\psi_{--}(\bm l,\bm l')|^2
Z(\bm l)Z(\bm l')
}{
\int d\Gamma(\bm k) 
|\psi_{--}(\bm k,\bm k)|^2
Z(\bm k)
+
(N-1)
\int d\Gamma(\bm k)  d\Gamma(\bm{k}')
|\psi_{--}(\bm k,\bm k')|^2
Z(\bm k)Z(\bm k')
}
}.
\end{eqnarray}
It should be stressed that also for the $\Psi_{22}(N)$ field operator the EPR average doesn't depend on the $N$ parameter
\begin{eqnarray}\label{epr22null}
&&
\frac{
\langle O(N)|
\Psi_{22}(N)^{\dagger}
Y_\beta (N)
Y_\alpha (N)
\Psi_{22}(N)
|O(N)\rangle
}
{
\langle O(N)|
\Psi_{22}(N)^{\dagger}
\Psi_{22}(N)
|O(N)\rangle}
\nonumber\\
&=&
\displaystyle{
\frac{2
\int_{\Omega} d\Gamma(\bm{l})~ 
\int_{\Omega'} d\Gamma(\bm{l}')~ 
\cos 2(\beta+\alpha-\theta_{22}(\bm l)-\theta_{22}(\bm l'))
|m_a(\bm l)m^a(\bm l')|^2
Z(\bm l)Z(\bm l')
}{
\int d\Gamma(\bm k)  d\Gamma(\bm{k}')
|m_a(\bm k)m^a(\bm k')|^2
Z(\bm k)Z(\bm k')
}
}.
\end{eqnarray}
Finally, using the relation between the polarization angles $\theta_{21}(\bm k)$ and  $\theta_{22}(\bm k)$ (\ref{rel21-22}), we see that
\begin{eqnarray}
\frac{
\langle O(N)|
\Psi_{21}(N)^{\dagger}
Y_\beta (N)
Y_\alpha (N)
\Psi_{21}(N)
|O(N)\rangle
}
{
\langle O(N)|
\Psi_{21}(N)^{\dagger}
\Psi_{21}(N)
|O(N)\rangle}
&=&
\frac{
\langle O(N)|
\Psi_{22}(N)^{\dagger}
Y_\beta (N)
Y_\alpha (N)
\Psi_{22}(N)
|O(N)\rangle
}
{
\langle O(N)|
\Psi_{22}(N)^{\dagger}
\Psi_{22}(N)
|O(N)\rangle}.
\end{eqnarray}
\\
From these calculations two main conclusion can be made. First one involves the $N$ parameter. In reducible representations the $N$ parameter does not necessary have to go to infinity, since each oscillator is a superposition of already infinitely many different momentum states. If we made an assumption that the $N$ parameter is a finite large number, it would have had an influence on the outcome of the EPR average for states maximally anti-correlated in circular basis. Like shown in section \ref{sec:detectionantycorr} the EPR average for maximally anti-correlated in circular polarization basis states depends on the $N$ parameter. The extra term in the denominator of the EPR averages for maximally anti-correlated in circular basis Bell states corresponding to the $\Psi_1(N)$ field operator may have influence on the outcome compared with the maximally correlated in circular basis Bell states corresponding to field operator $\Psi_2(N)$. Putting it another way, if any experiments confirmed a smaller outcome of the EPR average for maximally anti-correlated in circular basis states comparing with maximally correlated in circular basis states, it could have spoken in favor for the $N$ parameter being a finite number.
\\
\\
The second conclusion involves the polarization angle which for this representation is dependent on momentum. But first as an example let us take the EPR average for the correlated in circular polarization basis field operator $\Psi_{2}(N)$
\begin{eqnarray}
&&
\displaystyle{
\frac{2
\int_{\Omega} d\Gamma(\bm{l})~ 
\int_{\Omega'} d\Gamma(\bm{l}')~ 
\cos 2(\beta+\alpha-\theta_{22}(\bm l)-\theta_{22}(\bm l'))
|\psi_{--}(\bm l,\bm l')|^2
Z(\bm l)Z(\bm l')
}{
\int d\Gamma(\bm k)  d\Gamma(\bm{k}')
|\psi_{--}(\bm k,\bm k')|^2
Z(\bm k)Z(\bm k')
}
}.
\end{eqnarray}
Number 2 in the numerator may look suspicious at first, but it can be shown that this comes from the symmetry of the $|\psi_{--}(\bm k,\bm k')|^2Z(\bm k)Z(\bm k')$ term. Denoting $f(\bm k, \bm k')=|\psi_{--}(\bm k,\bm k')|^2Z(\bm k)Z(\bm k')$, we see that $f(\bm k, \bm k')=f(\bm k', \bm k)$, and
\begin{eqnarray}
&&
2
\int\limits_{\Omega} d\Gamma(\bm{l})~ 
\int\limits_{\Omega'} d\Gamma(\bm{l}')~ 
\cos 2(\beta+\alpha-\theta_{22}(\bm l)-\theta_{22}(\bm l')) f(\bm l,\bm l')
\nonumber\\
&=&
\int\limits_{\Omega} d\Gamma(\bm{l})~ 
\int\limits_{\Omega'} d\Gamma(\bm{l}')~ 
\cos 2(\beta+\alpha-\theta_{22}(\bm l)-\theta_{22}(\bm l')) f(\bm l,\bm l')
\nonumber\\
&+&
\int\limits_{\Omega} d\Gamma(\bm{l})~ 
\int\limits_{\Omega'} d\Gamma(\bm{l}')~ 
\cos 2(\beta+\alpha-\theta_{22}(\bm l')-\theta_{22}(\bm l)) f(\bm l',\bm l)
\nonumber\\
&=&
\displaystyle\int_{(\Omega \times \Omega') \cup (\Omega' \times \Omega)} d\Gamma(\bm{l})~ 
 d\Gamma(\bm{l}')~ 
\cos 2(\beta+\alpha-\theta_{22}(\bm l)-\theta_{22}(\bm l'))
f(\bm l,\bm l').
\end{eqnarray}
We find that $(\Omega \times \Omega') \cup (\Omega' \times \Omega) \subset R^3\times R^3$, $f(\bm k, \bm k')$ is always nonnegative and the cosine term is bounded, i.e. $|\cos 2(\beta+\alpha-\theta_{22}(\bm l)-\theta_{22}(\bm l'))|\le 1 $, which implies that
\begin{eqnarray}\label{EPRle1}
&&
\frac{
\int_{(\Omega \times \Omega') \cup (\Omega' \times \Omega)} d\Gamma(\bm{l})~ 
 d\Gamma(\bm{l}')~ 
\cos 2(\beta+\alpha-\theta_{22}(\bm l)-\theta_{22}(\bm l'))
f(\bm l,\bm l')
}
{
\int_{R^3\times R^3} d\Gamma(\bm k)  d\Gamma(\bm{k}')
f(\bm k,\bm k')
}\le 1.
\end{eqnarray}
As we can see the EPR average in such reducible representation with the polarization angle dependent on momentum may serve a shift of phase comparing with ``standard theory models". Other than that, for the vacuum probability density function $Z(\bm k)$ being flat in the detector's momentum solid angle spread, is hard to distinguish this result from ``standard models".
\\
\section {EPR-type experiment under Lorentz transformation}\label{sec:eprtrans}
Now the relativistic correlations for photon Bell states in the background of reducible representations of $N$-oscillator algebras will be considered. For the relativistic considerations the field should be a Lorentz covariant one, and the invariance of the four Bell state corresponding field operators of the proposed model was shown in previous section \ref{sec:trans}.
Let us then consider two observers moving relative to each other with detectors that detect an EPR pair. As far as simultaneity of events is considered the moment of the collapse of the field will depend on the reference frame and observers may have a disagreement on whose detectors clicked first. Therefore for considerations made in this section the simultaneity problem is not an issue in EPR experiment, where in the experiment we ask a question about the correlations. In other words, although observers may not agree on whose detector clicked first, they will agree on the outcome of the experiment, i.e. the correlations. In this sense there in no preferred frame of reference.
\subsection{Relativistic correlation of a two-photon state - case 1}\label{sec:eprtranscase1}
Within the framework of relativistic quantum field theory, let us consider the Einstein-Podolsky-Rosen (EPR) gedankenexperiment in which
measurements on detectors are performed by moving observers. In this section we perform a Lorenz transformation on both detectors Alice's and Bob's, so that Alice and Bob are in the same reference frame, not moving with respect to each other. We will start with an arbitrary two-photon state.
\\
First let us notice that in detectors modeled by a yes-no observable (\ref{YNcirc}) after a Lorentz transformation the $\alpha$ orientation angle is observed to be shifted by the Wigner phase $2\Theta(\Lambda,\bm{r})$.  
\begin{eqnarray}\label{transY}
U(\Lambda,N)^{\dagger}
Y_\alpha (\bm l, N)
U(\Lambda,N)
&=&
\sum_{n=1}^{N}
\sum_{s=\pm}
\left(
e^{2is\alpha}
e^{-4is\Theta(\Lambda,\bm{l})}
|\bm{\Lambda^{-1}l}
\rangle\langle 
\bm{\Lambda^{-1}l}|
\otimes
a_{-s}^{\dagger}
a_{s}
\right)^{(n)}
.
\end{eqnarray}
So that under Lorentz transformation performed on both detectors we have an unnormalized EPR average of the form
\begin{eqnarray}\label{scalarPsitransYYPsi}
&&
\langle O(N)|
\Psi(N)^{\dagger}
U(\Lambda,N)^{\dagger}
Y_\beta (N)
Y_\alpha (N)
U(\Lambda,N)
\Psi(N)
|O(N)\rangle
\nonumber\\
&=&
4
\sum_{ss'=\pm}
\int\limits_{\Omega} d\Gamma(\bm{l})~ 
\int\limits_{\Omega'} d\Gamma(\bm{l}')~ 
e^{-2i(s\beta+s'\alpha)}
e^{4i(s\Theta(\Lambda,\bm{l})+s'\Theta(\Lambda,\bm{l}'))}
\bar{\psi}_{ss'}(\bm{\Lambda^{-1} l},\bm{\Lambda^{-1} l}')
\psi_{-s'-s}(\bm{\Lambda^{-1} l}',\bm{\Lambda^{-1} l})
\nonumber\\
&\times&
\langle O(N)|
I(\bm{\Lambda^{-1} l},N)I(\bm{\Lambda^{-1} l}',N)
|O(N)\rangle
\nonumber\\
&+&
4
\sum_{ss'=\pm}
\int\limits_{\Omega} d\Gamma(\bm{l})~ 
\int\limits_{\Omega'} d\Gamma(\bm{l}')~ 
\int d\Gamma(\bm k)  ~ 
e^{-2i(s\beta-s\alpha)}
e^{4i(s\Theta(\Lambda,\bm{l})-s\Theta(\Lambda,\bm{l}'))}
\nonumber\\
&\times&
\bar{\psi}_{ss'}(\bm{\Lambda^{-1} l},\bm k)
\psi_{ss'}(\bm{\Lambda^{-1} l}',\bm k)
\delta_{\Gamma}(\bm{\Lambda^{-1} l}, \bm{\Lambda^{-1} l}')
\langle O(N)|
I(\bm k,N)I(\bm{\Lambda^{-1} l},N)
|O(N)\rangle.
\end{eqnarray}
In the case of disjoint detectors just one part has contribution. Also from the transformation rule on the fields (\ref{Psiss'trans}) we get:
\begin{eqnarray}
&&
\langle O(N)|
\Psi(N)^{\dagger}
U(\Lambda,N)^{\dagger}
Y_\beta (N)
Y_\alpha (N)
U(\Lambda,N)
\Psi(N)
|O(N)\rangle
\nonumber\\
&=&
\frac{4(N-1)}{N}
\sum_{ss'=\pm}
\int\limits_{\Lambda \Omega} d\Gamma(\bm{l})~ 
\int\limits_{\Lambda \Omega'} d\Gamma(\bm{l}')~ 
e^{-2i(s\beta+s'\alpha)}
e^{4i(s\Theta(\Lambda,\bm{\Lambda l})+s'\Theta(\Lambda,\bm{\Lambda l}'))}
\bar{\psi}_{ss'}(\bm l,\bm l')
\psi_{-s-s'}(\bm l,\bm l')
Z(\bm{ l})Z(\bm{l}').
\nonumber\\
\end{eqnarray}
So the unnormalized EPR average, in the case when both detectors undergo the same Lorentz transformation, for a two-photon state can be written in the form 
\begin{eqnarray}
&&
\langle O(N)|
\Psi(N)^{\dagger}
U(\Lambda,N)^{\dagger}
Y_\beta (N)
Y_\alpha (N)
U(\Lambda,N)
\Psi(N)
|O(N)\rangle
\nonumber\\
&=&
\frac{8(N-1)}{N}
\int\limits_{\Lambda \Omega} d\Gamma(\bm{l})~ 
\int\limits_{\Lambda \Omega'} d\Gamma(\bm{l}')~ 
 \cos 2(\beta+\alpha
-2\Theta(\Lambda,\bm{\Lambda l})-2\Theta(\Lambda,\bm{\Lambda l}'))
\Re\left(
\bar{\psi}_{++}(\bm l,\bm l')
\psi_{--}(\bm l,\bm l')
\right)
Z(\bm{ l})Z(\bm{l}')
\nonumber\\
&+&
\frac{8(N-1)}{N}
\int\limits_{\Lambda \Omega} d\Gamma(\bm{l})~ 
\int\limits_{\Lambda \Omega'} d\Gamma(\bm{l}')~ 
\sin 2(\beta+\alpha-2\Theta(\Lambda,\bm{\Lambda l})-2\Theta(\Lambda,\bm{\Lambda l}'))
\Im\left(
\bar{\psi}_{++}(\bm l,\bm l')
\psi_{--}(\bm l,\bm l')
\right)
Z(\bm{ l})Z(\bm{l}')
\nonumber\\
&+&
\frac{8(N-1)}{N}
\int\limits_{\Lambda \Omega} d\Gamma(\bm{l})~ 
\int\limits_{\Lambda \Omega'} d\Gamma(\bm{l}')~ 
\cos 2(\beta-\alpha-2\Theta(\Lambda,\bm{\Lambda l})+2\Theta(\Lambda,\bm{\Lambda l}'))
\Re\left(
\bar{\psi}_{+-}(\bm l,\bm l')
\psi_{-+}(\bm l,\bm l')
\right)
Z(\bm{ l})Z(\bm{l}')
\nonumber\\
&+&
\frac{8(N-1)}{N}
\int\limits_{\Lambda \Omega} d\Gamma(\bm{l})~ 
\int\limits_{\Lambda \Omega'} d\Gamma(\bm{l}')~ 
\sin 2(\beta-\alpha-2\Theta(\Lambda,\bm{\Lambda l})+2\Theta(\Lambda,\bm{\Lambda l}'))
\Im\left(
\bar{\psi}_{+-}(\bm l,\bm l')
\psi_{-+}(\bm l,\bm l')
\right)
Z(\bm{ l})Z(\bm{l}').
\nonumber\\
\end{eqnarray}
\subsection{Relativistic correlation function for maximally anti-correlated in circular polarization basis states - case 1}\label{sec:eprtranscase1anty}
Now let us consider such relativistic correlation function for the Bell states. We will start from the maximally anti-correlated states in circular polarization basis corresponding to the field operator $\Psi_1(N)$ (\ref{Psi1}). In a realistic case, when the localization of the photon detectors leads to  momentum solid angle spreads 
$\bm l\in \Omega, ~\bm l'\in \Omega'$ and the detectors are disjoint $\Omega \cap \Omega' = \emptyset$, the correlation function reads:
~~
\begin{eqnarray}
&&
\langle O(N)|
\Psi_{1}(N)^{\dagger}
U(\Lambda,N)^{\dagger}
Y_\beta (N)
Y_\alpha (N)
U(\Lambda,N)
\Psi_{1}(N)
|O(N)\rangle
\nonumber\\
&=&
\frac{8(N-1)}{N}
\int\limits_{\Lambda \Omega} d\Gamma(\bm{l})~ 
\int\limits_{\Lambda \Omega'} d\Gamma(\bm{l}')~ 
\cos 2(\beta-\alpha-2\Theta(\Lambda,\bm{\Lambda l})+2\Theta(\Lambda,\bm{\Lambda l}'))
\Re\left(
\bar{\psi}_{+-}(\bm l,\bm l')
\psi_{-+}(\bm l,\bm l')
\right)
Z(\bm{ l})Z(\bm{l}')
\nonumber\\
&+&
\frac{8(N-1)}{N}
\int\limits_{\Lambda \Omega} d\Gamma(\bm{l})~ 
\int\limits_{\Lambda \Omega'} d\Gamma(\bm{l}')~ 
\sin 2(\beta-\alpha-2\Theta(\Lambda,\bm{\Lambda l})+2\Theta(\Lambda,\bm{\Lambda l}'))
\Im\left(
\bar{\psi}_{+-}(\bm l,\bm l')
\psi_{-+}(\bm l,\bm l')
\right)
Z(\bm{ l})Z(\bm{l}')
.
\nonumber\\
\end{eqnarray}
For the field operator $\Psi_{11}(N)$ (\ref{Psi11}), using the condition on the field and the polarization angle (\ref{bellcond11}), we get
\begin{eqnarray}
&&
\langle O(N)|
\Psi_{11}(N)^{\dagger}
U(\Lambda,N)^{\dagger}
Y_\beta (N)
Y_\alpha (N)
U(\Lambda,N)
\Psi_{11}(N)
|O(N)\rangle
\nonumber\\
&=&
\frac{-8(N-1)}{N}
\int\limits_{\Lambda \Omega} d\Gamma(\bm{l})~ 
\int\limits_{\Lambda \Omega'} d\Gamma(\bm{l}')~ 
\cos 2(\beta-\alpha
-\theta_{11}(\bm{\Lambda l})+\theta_{11}(\bm{\Lambda l}'))
|\psi_{-+}(\bm l,\bm l')|^2
Z(\bm{ l})Z(\bm{l}')
\label{Psi11redet}
\\
&=&
\frac{-8(N-1)}{N}\label{Psi11revac}
\int\limits_{\Omega} d\Gamma(\bm{l})~ 
\int\limits_{\Omega'} d\Gamma(\bm{l}')~ 
\cos 2(\beta-\alpha
-\theta_{11}(\bm{ l})+\theta_{11}(\bm{ l}'))
|\psi_{-+}(\bm l,\bm l')|^2
Z(\bm{\Lambda^{-1 }l})Z(\bm{\Lambda^{-1}l}').
\end{eqnarray}
Comparing this to (\ref{epr11}) we find that the vacuum probability density may have an influence on the correlation of the detectors outcome.\\
\\
For the Bell state corresponding to the $\Psi_{12}(N)$ field operator (\ref{Psi12}), using the condition on the field and the polarization angle (\ref{bellcond12}), we get
\begin{eqnarray}
&&
\langle O(N)|
\Psi_{12}(N)^{\dagger}
U(\Lambda,N)^{\dagger}
Y_\beta (N)
Y_\alpha (N)
U(\Lambda,N)
\Psi_{12}(N)
|O(N)\rangle
\nonumber\\
&=&
\frac{8(N-1)}{N}
\int\limits_{\Lambda \Omega} d\Gamma(\bm{l})~ 
\int\limits_{\Lambda \Omega'} d\Gamma(\bm{l}')~ 
\cos 2(\beta-\alpha-\theta_{12}(\bm{\Lambda l})+\theta_{12}(\bm{\Lambda l}'))
|\psi_{-+}(\bm l,\bm l')|^2
Z(\bm{ l})Z(\bm{l}')
\label{Psi12redet}
\\
&=&
\frac{8(N-1)}{N}
\int\limits_{\Omega} d\Gamma(\bm{l})~ 
\int\limits_{\Omega'} d\Gamma(\bm{l}')~ 
\cos 2(\beta-\alpha-\theta_{12}(\bm{l})+\theta_{12}(\bm{l}'))
|\psi_{-+}(\bm l,\bm l')|^2
Z(\bm{ \Lambda^{-1}l})Z(\bm{\Lambda^{-1}l}').
\label{Psi12revac}
\end{eqnarray}
One may look at these formulas from two points of view: as a transformation on the detector angle spreads and the polarization angles  (\ref{Psi11redet}), (\ref{Psi12redet}) or a transformation on the vacuum probability density function (\ref{Psi11revac}), (\ref{Psi12revac}).
\subsection{Relativistic correlation function for maximally correlated in circular polarization basis states - case 1}\label{sec:eprtranscase1corr}
For maximally correlated in circular polarization basis field operators let us consider the $\Psi_2(N)$ field operator (\ref{Psi2}).
In a realistic case, when the localization of the photon 
detectors leads to momentum solid angle spreads 
$\bm l\in \Omega, ~\bm l'\in \Omega'$ and for disjoint detectors 
$\Omega \cap \Omega' = \emptyset$, the correlation function reads:
\begin{eqnarray}
&&
\langle O(N)|
\Psi_2(N)^{\dagger}
U(\Lambda,N)^{\dagger}
Y_\beta (N)
Y_\alpha (N)
U(\Lambda,N)
\Psi_2(N)
|O(N)\rangle
\nonumber\\
&=&
\frac{8(N-1)}{N}
\int\limits_{\Lambda\Omega} d\Gamma(\bm{l})~ 
\int\limits_{\Lambda\Omega'} d\Gamma(\bm{l}')~ 
\cos 2(\beta+\alpha
-2\Theta(\Lambda,\bm{\Lambda l})-2\Theta(\Lambda,\bm{\Lambda l}'))
\Re\left(
\bar{\psi}_{++}(\bm l,\bm l')
\psi_{--}(\bm l,\bm l')
\right)
Z(\bm l)Z(\bm l')
\nonumber\\
&+&
\frac{8(N-1)}{N}
\int\limits_{\Lambda\Omega} d\Gamma(\bm{l})~ 
\int\limits_{\Lambda\Omega'} d\Gamma(\bm{l}')~ 
\sin 2(\beta+\alpha
-2\Theta(\Lambda,\bm{\Lambda l})-2\Theta(\Lambda,\bm{\Lambda l}'))
\Im\left(
\bar{\psi}_{++}(\bm l,\bm l')
\psi_{--}(\bm l,\bm l')
\right)
Z(\bm l)Z(\bm l')
.
\nonumber\\
\end{eqnarray}
For the Bell state corresponding to the $\Psi_{12}(N)$ field operator (\ref{Psi21}) using the condition on the field and the polarization angle (\ref{bellcond21}) we get the following unnormalized relativistic EPR average
\begin{eqnarray}
&&
\langle O(N)|
\Psi_{21}(N)^{\dagger}
U(\Lambda,N)^{\dagger}
Y_\beta (N)
Y_\alpha (N)
U(\Lambda,N)
\Psi_{21}(N)
|O(N)\rangle
\nonumber\\
&=&
\frac{-8(N-1)}{N}
\int\limits_{\Lambda \Omega} d\Gamma(\bm{l})~ 
\int\limits_{\Lambda \Omega'} d\Gamma(\bm{l}')~ 
\cos 2(\beta+\alpha-\theta_{21}(\bm{\Lambda l})-\theta_{21}(\bm{\Lambda l}'))
|\psi_{--}(\bm l,\bm l')|^2
Z(\bm{ l})Z(\bm{l}')
\\
&=&
\frac{-8(N-1)}{N}
\int\limits_{\Omega} d\Gamma(\bm{l})~ 
\int\limits_{\Omega'} d\Gamma(\bm{l}')~ 
\cos 2(\beta+\alpha-\theta_{21}(\bm{l})-\theta_{21}(\bm{l}'))
|\psi_{--}(\bm l,\bm l')|^2
Z(\bm{\Lambda^{-1} l})Z(\bm{\Lambda^{-1}l}').
\end{eqnarray}
Finally for the Bell state corresponding to the $\Psi_{22}(N)$ field operator (\ref{Psi22}), using the condition on the field and the polarization function (\ref{bellcond22}), we get
\begin{eqnarray}
&&
\langle O(N)|
\Psi_{22}(N)^{\dagger}
U(\Lambda,N)^{\dagger}
Y_\beta (N)
Y_\alpha (N)
U(\Lambda,N)
\Psi_{22}(N)
|O(N)\rangle
\nonumber\\
&=&
\frac{8(N-1)}{N}
\int\limits_{\Lambda \Omega} d\Gamma(\bm{l})~ 
\int\limits_{\Lambda \Omega'} d\Gamma(\bm{l}')~ 
\cos 2(\beta+\alpha-\theta_{22}(\bm{\Lambda l})-\theta_{22}(\bm{\Lambda l}'))
|\psi_{--}(\bm l,\bm l')|^2
Z(\bm{ l})Z(\bm{l}')
\\
&=&
\frac{8(N-1)}{N}
\int\limits_{\Omega} d\Gamma(\bm{l})~ 
\int\limits_{\Omega'} d\Gamma(\bm{l}')~ 
\cos 2(\beta+\alpha-\theta_{22}(\bm{l})-\theta_{22}(\bm{l}'))
|\psi_{--}(\bm l,\bm l')|^2
Z(\bm{\Lambda^{-1} l})Z(\bm{\Lambda^{-1} l}').
\end{eqnarray}
Again we may look at these formulas from two points of view: as a transformation on the detectors angle spread and the polarization angle or a transformation on the vacuum probability density.
\subsection{Relativistic correlation of a two-photon state - case 2}\label{sec:eprtranscase2}
Now within the framework of relativistic quantum field theory, let us consider the Einstein-Podolsky-Rosen (EPR) gedankenexperiment in which
measurements on detectors are performed by moving observers, only this time we perform a Lorentz transformation only on Alice's detector, so that both detectors are moving with respect to each other. 
\\
Let us first consider the field operator corresponding to an arbitrary two-photon state (\ref{PsiNs}). Under Lorentz transformation on just one detector we have
\begin{eqnarray}
&&
\langle O(N)|
\Psi(N)^{\dagger}
Y_\beta (N)
U(\Lambda,N)^{\dagger}
Y_\alpha (N)
U(\Lambda,N)
\Psi(N)
|O(N)\rangle
\nonumber\\
&=&
4
\sum_{ss'=\pm}
\int\limits_{\Omega} d\Gamma(\bm{l})~ 
\int\limits_{\Omega'} d\Gamma(\bm{l}')~ 
e^{-2i(s\beta+s'\alpha)}
e^{4is'\Theta(\Lambda,\bm{l}')}
\bar{\psi}_{ss'}(\bm{l},\bm{\Lambda^{-1} l}')
\psi_{-s-s'}(\bm{l},\bm{\Lambda^{-1} l}')
\nonumber\\
&\times&
\langle O(N)|
I(\bm{l},N)I(\bm{\Lambda^{-1} l}',N)
|O(N)\rangle
\nonumber\\
&+&
4
\sum_{ss'=\pm}
\int\limits_{\Omega} d\Gamma(\bm{l})~ 
\int\limits_{\Omega'} d\Gamma(\bm{l}')~ 
\int d\Gamma(\bm k)  ~ 
e^{-2i(s\beta-s\alpha)}
e^{4is\Theta(\Lambda,\bm{l}')}
\bar{\psi}_{ss'}(\bm{l},\bm k)
\psi_{ss'}(\bm{\Lambda^{-1} l}',\bm k)
\delta_{\Gamma}(\bm{l}, \bm{\Lambda^{-1} l}')
\nonumber\\
&\times&
\langle O(N)|
I(\bm k,N)I(\bm{ l},N)
|O(N)\rangle.
\end{eqnarray}
In the case of disjoint detectors just one part has contribution. Also from the transformation rule on the fields (\ref{Psiss'trans}) we get:
\begin{eqnarray}\label{scalarPsiYtransYPsi}
&&
\langle O(N)|
\Psi(N)^{\dagger}
Y_\beta (N)
U(\Lambda,N)^{\dagger}
Y_\alpha (N)
U(\Lambda,N)
\Psi(N)
|O(N)\rangle
\nonumber\\
&=&
\frac{4(N-1)}{N}
\sum_{ss'=\pm}
\int\limits_{\Omega} d\Gamma(\bm{l})~ 
\int\limits_{\Lambda \Omega'} d\Gamma(\bm{l}')~ 
e^{-2i(s\beta+s'\alpha)}
e^{4i(s'\Theta(\Lambda,\bm{\Lambda l}'))}
\bar{\psi}_{ss'}(\bm l,\bm l')
\psi_{-s-s'}(\bm l,\bm l')
Z(\bm{ l})Z(\bm{l}').
\end{eqnarray}
Then the unnormalized relativistic EPR average for a two-photon state, in a situation when a Lorentz transformation is performed only on Alice's detector, can be written in the form
\begin{eqnarray}
&&
\langle O(N)|
\Psi(N)^{\dagger}
Y_\beta (N)
U(\Lambda,N)^{\dagger}
Y_\alpha (N)
U(\Lambda,N)
\Psi(N)
|O(N)\rangle
\nonumber\\
&=&
\frac{8(N-1)}{N}
\int\limits_{\Omega} d\Gamma(\bm{l})~ 
\int\limits_{\Lambda \Omega'} d\Gamma(\bm{l}')~ 
 \cos 2(\beta+\alpha
-2\Theta(\Lambda,\bm{\Lambda l}'))
\Re\left(
\bar{\psi}_{++}(\bm l,\bm l')
\psi_{--}(\bm l,\bm l')
\right)
Z(\bm{ l})Z(\bm{l}')
\nonumber\\
&+&
\frac{8(N-1)}{N}
\int\limits_{\Omega} d\Gamma(\bm{l})~ 
\int\limits_{\Lambda \Omega'} d\Gamma(\bm{l}')~ 
\sin 2(\beta+\alpha-2\Theta(\Lambda,\bm{\Lambda l}'))
\Im\left(
\bar{\psi}_{++}(\bm l,\bm l')
\psi_{--}(\bm l,\bm l')
\right)
Z(\bm{ l})Z(\bm{l}')
\nonumber\\
&+&
\frac{8(N-1)}{N}
\int\limits_{\Omega} d\Gamma(\bm{l})~ 
\int\limits_{\Lambda \Omega'} d\Gamma(\bm{l}')~ 
\cos 2(\beta-\alpha+2\Theta(\Lambda,\bm{\Lambda l}'))
\Re\left(
\bar{\psi}_{+-}(\bm l,\bm l')
\psi_{-+}(\bm l,\bm l')
\right)
Z(\bm{ l})Z(\bm{l}')
\nonumber\\
&+&
\frac{8(N-1)}{N}
\int\limits_{\Omega} d\Gamma(\bm{l})~ 
\int\limits_{\Lambda \Omega'} d\Gamma(\bm{l}')~ 
\sin 2(\beta-\alpha+2\Theta(\Lambda,\bm{\Lambda l}'))
\Im\left(
\bar{\psi}_{+-}(\bm l,\bm l')
\psi_{-+}(\bm l,\bm l')
\right)
Z(\bm{ l})Z(\bm{l}').
\end{eqnarray}
\subsection{Relativistic correlation function for maximally anti-correlated in circular polarization basis states - case 2}\label{sec:eprtranscase2anty}
In a realistic case, when the localization of the photon 
detectors leads to  momentum solid angle spreads 
$\bm l\in \Omega, ~\bm l'\in \Omega'$ and for disjoint detectors 
$\Omega \cap \Omega' = \emptyset$, the correlation function reads:
\begin{eqnarray}
&&
\langle O(N)|
\Psi_{1}(N)^{\dagger}
Y_\beta (N)
U(\Lambda,N)^{\dagger}
Y_\alpha (N)
U(\Lambda,N)
\Psi_{1}(N)
|O(N)\rangle
\nonumber\\
&=&
\frac{8(N-1)}{N}
\int\limits_{\Omega} d\Gamma(\bm{l})~ 
\int\limits_{\Lambda \Omega'} d\Gamma(\bm{l}')~ 
\cos 2(\beta-\alpha+2\Theta(\Lambda,\bm{\Lambda l}'))
\Re\left(
\bar{\psi}_{+-}(\bm l,\bm l')
\psi_{-+}(\bm l,\bm l')
\right)
Z(\bm{ l})Z(\bm{l}')
\nonumber\\
&+&
\frac{8(N-1)}{N}
\int\limits_{\Omega} d\Gamma(\bm{l})~ 
\int\limits_{\Lambda \Omega'} d\Gamma(\bm{l}')~ 
\sin 2(\beta-\alpha+2\Theta(\Lambda,\bm{\Lambda l}'))
\Im\left(
\bar{\psi}_{+-}(\bm l,\bm l')
\psi_{-+}(\bm l,\bm l')
\right)
Z(\bm{l})Z(\bm{l}').
\end{eqnarray}
For the Bell state corresponding to the $\Psi_{11}(N)$ field operator (\ref{Psi11}), using condition (\ref{bellcond11}), we get an unnormalized relativistic EPR average of the form
\begin{eqnarray}
&&
\langle O(N)|
\Psi_{11}(N)^{\dagger}
Y_\beta (N)
U(\Lambda,N)^{\dagger}
Y_\alpha (N)
U(\Lambda,N)
\Psi_{11}(N)
|O(N)\rangle
\nonumber\\
&=&
\frac{-8(N-1)}{N}
\int\limits_{\Omega} d\Gamma(\bm{l})~ 
\int\limits_{\Lambda \Omega'} d\Gamma(\bm{l}')~ 
\cos 2(\beta-\alpha-\theta_{11}(\bm{l})+\theta_{11}(\bm{\Lambda l}'))
|\psi_{-+}(\bm l,\bm l')|^2
Z(\bm{ l})Z(\bm{l}')\label{Psi11Aredet}
\\
&=&
\frac{-8(N-1)}{N}
\int\limits_{\Lambda^{-1}\Omega} d\Gamma(\bm{l})~ 
\int\limits_{\Omega'} d\Gamma(\bm{l}')~ 
\cos 2(\beta-\alpha-\theta_{11}(\bm{\Lambda^{-1} l} )+\theta_{11}(\bm{ l}'))
|\psi_{-+}(\bm l,\bm l')|^2
Z(\bm{\Lambda^{-1} l })Z(\bm{\Lambda^{-1}l}').\label{Psi11Bredet}
\nonumber\\
\end{eqnarray}
Next for the Bell state corresponding to the $\Psi_{12}(N)$  field operator (\ref{Psi12}), and using condition (\ref{bellcond12}), we get
\begin{eqnarray}
&&
\langle O(N)|
\Psi_{12}(N)^{\dagger}
Y_\beta (N)
U(\Lambda,N)^{\dagger}
Y_\alpha (N)
U(\Lambda,N)
\Psi_{12}(N)
|O(N)\rangle
\nonumber\\
&=&
\frac{8(N-1)}{N}
\int\limits_{\Omega} d\Gamma(\bm{l})~ 
\int\limits_{\Lambda \Omega'} d\Gamma(\bm{l}')~ 
\cos 2(\beta-\alpha
-\theta_{12}(\bm{ l})+\theta_{12}(\bm{\Lambda l}'))
|\psi_{-+}(\bm l,\bm l')|^2
Z(\bm{l})Z(\bm{l}')\label{Psi12Aredet}
\\
&=&
\frac{8(N-1)}{N}
\int\limits_{\Lambda^{-1}\Omega } d\Gamma(\bm{l})~ 
\int\limits_{\Omega'} d\Gamma(\bm{l}')~ 
\cos 2(\beta-\alpha-\theta_{12}(\bm{\Lambda^{-1}l })+\theta_{12}(\bm{l}'))
|\psi_{-+}(\bm l,\bm l')|^2
Z(\bm{\Lambda^{-1}l })Z(\bm{\Lambda^{-1}l}').\label{Psi12Bredet}
\nonumber\\
\end{eqnarray}
\subsection{Relativistic correlation function for maximally correlated in circular polarization basis states - case 2}\label{sec:eprtranscase2corr}
For maximally correlated in circular polarization basis field operators let us consider $\Psi_2(N)$.
In a realistic case, when the localization of the photon 
detectors leads to momentum solid angle spreads 
$\bm{l}\in \Omega, ~\bm{l}'\in \Omega'$ and for disjoint detectors 
$\Omega \cap \Omega' = \emptyset$, the correlation function reads:
\begin{eqnarray}
&&
\langle O(N)|
\Psi_2(N)^{\dagger}
Y_\beta (N)
U(\Lambda,N)^{\dagger}
Y_\alpha (N)
U(\Lambda,N)
\Psi_2(N)
|O(N)\rangle
\nonumber\\
&=&
\frac{8(N-1)}{N}
\int\limits_{\Omega} d\Gamma(\bm{l})~ 
\int\limits_{\Lambda\Omega'} d\Gamma(\bm{l}')~ 
\cos 2(\beta+\alpha
-2\Theta(\Lambda,\bm{\Lambda l}'))
\Re\left(
\bar{\psi}_{++}(\bm l,\bm l')
\psi_{--}(\bm l,\bm l')
\right)
Z(\bm l)Z(\bm l')
\nonumber\\
&+&
\frac{8(N-1)}{N}
\int\limits_{\Omega} d\Gamma(\bm{l})~ 
\int\limits_{\Lambda\Omega'} d\Gamma(\bm{l}')~ 
\sin 2(\beta+\alpha
-2\Theta(\Lambda,\bm{\Lambda l}'))
\Im\left(
\bar{\psi}_{++}(\bm l,\bm l')
\psi_{--}(\bm l,\bm l')
\right)
Z(\bm l)Z(\bm l')
.
\end{eqnarray}
Now for the Bell state corresponding to the $\Psi_{21}(N)$ field operator (\ref{Psi21}), using the condition (\ref{bellcond21}), we get the following unnormalized EPR average
\begin{eqnarray}
&&
\langle O(N)|
\Psi_{21}(N)^{\dagger}
Y_\beta (N)
U(\Lambda,N)^{\dagger}
Y_\alpha (N)
U(\Lambda,N)
\Psi_{21}(N)
|O(N)\rangle
\nonumber\\
&=&
\frac{-8(N-1)}{N}
\int\limits_{\Omega} d\Gamma(\bm{l})~ 
\int\limits_{\Lambda \Omega'} d\Gamma(\bm{l}')~ 
\cos 2(\beta+\alpha-\theta_{21}(\bm{l})-\theta_{21}(\bm{\Lambda l}'))
|\psi_{--}(\bm l,\bm l')|^2
Z(\bm{ l})Z(\bm{l}')\label{Psi21Aredet}
\\
&=&
\frac{-8(N-1)}{N}
\int\limits_{\Lambda^{-1}\Omega} d\Gamma(\bm{l})~ 
\int\limits_{\Omega'} d\Gamma(\bm{l}')~ 
\cos 2(\beta+\alpha-\theta_{21}(\bm{\Lambda^{-1}l })-\theta_{21}(\bm{l}'))
|\psi_{--}(\bm l,\bm l')|^2
Z(\bm{\Lambda^{-1}l })Z(\bm{\Lambda^{-1}l}').\label{Psi21Bredet}
\nonumber\\
\end{eqnarray}
Finally for the Bell state corresponding to the $\Psi_{22}(N)$ field operator (\ref{Psi22}), using the condition (\ref{bellcond22}), we get
\begin{eqnarray}
&&
\langle O(N)|
\Psi_{22}(N)^{\dagger}
Y_\beta (N)
U(\Lambda,N)^{\dagger}
Y_\alpha (N)
U(\Lambda,N)
\Psi_{22}(N)
|O(N)\rangle
\nonumber\\
&=&
\frac{8(N-1)}{N}
\int\limits_{\Omega} d\Gamma(\bm{l})~ 
\int\limits_{\Lambda \Omega'} d\Gamma(\bm{l}')~ 
\cos 2(\beta+\alpha-\theta_{22}(\bm{ l})-\theta_{22}(\bm{\Lambda l}'))
|\psi_{--}(\bm l,\bm l')|^2
Z(\bm{ l})Z(\bm{l}')\label{Psi22Aredet}
\\
&=&
\frac{8(N-1)}{N}
\int\limits_{\Lambda^{-1}\Omega} d\Gamma(\bm{l})~ 
\int\limits_{\Omega'} d\Gamma(\bm{l}')~ 
\cos 2(\beta+\alpha-\theta_{22}(\bm{\Lambda^{-1}l})-\theta_{22}(\bm{l}'))
|\psi_{--}(\bm l,\bm l')|^2
Z(\bm{\Lambda^{-1} l})Z(\bm{\Lambda^{-1} l}').\label{Psi22Bredet}
\nonumber\\
\end{eqnarray}
We may look at these formulas from two points of view: as a transformation on the Alice's detector (\ref{Psi11Aredet}), (\ref{Psi12Aredet}), (\ref{Psi21Aredet}) and (\ref{Psi22Aredet}) or an inverse transformation on Bob's detector with a compensating Lorentz transformation on the vacuum probability density (\ref{Psi11Bredet}), (\ref{Psi12Bredet}), (\ref{Psi21Bredet}) and (\ref{Psi22Bredet}).
\\
\section{Summary and conclusions}
The main motivation for this work was to take a closer look at a relativistic model for photon fields in reducible representations of harmonic oscillator Lie algebras (HOLA) proposed by Czachor \cite{MC00}-\cite{MCLecture} with an application to relativistic EPR-type correlations. 
On the other hand, employing reducible representations for relativistic EPR-type experiments may show the role that the oscillator number $N$ and vacuum probability density $Z(\bm k)$, known from such representations, play in this model. \\\\
It has been shown that it is possible to model Bell states in quantum field theory background of $N$-oscillator reducible representations. The main assumption is that Bell states are maximally correlated or maximally anti-correlated in two polarization bases: circular and linear. However it should be stressed here that in this model the linear polarization angles are dependent on momentum, and from the condition for maximal correlation in both bases we get conditions on the fields and polarization angle functions (\ref{bellcond11}), (\ref{bellcond12}), (\ref{bellcond21}) and (\ref{bellcond22}). It turns out that employing such momentum dependent polarization angle is important for maintaining Lorentz covariance in both bases. \\
\\
Further it has been shown that theoretically it is possible to maintain Lorentz covariance of the field operators corresponding to the photon Bell states introduced earlier in both polarization basis. The conclusion is: to obtain maximal correlation for EPR-type experiments in both bases one has to employ momentum dependent polarization functions that transform under Lorentz transformation in such a way that they compensate the Wigner phase $2\Theta(\Lambda,\bm{k})$.
\\\\
Next the EPR correlation functions for all of the four Bell states where calculated in reducible representations. First conclusion from this involves the $N$ parameter. In reducible representations  the $N$ parameter does not necessary have to go to infinity, since each oscillator is a superposition of already infinitely many different momentum states. If we made an assumption that the $N$ parameter is a finite large number, it would have had an influence on the outcome of the EPR average for states maximally anti-correlated in circular basis. Like shown in section \ref{sec:detectionantycorr} the EPR average for maximally anti-correlated in circular polarization basis states depends on the $N$ parameter. The extra term in the denominator of the EPR averages for Bell states corresponding to the $\Psi_1(N)$ field operator may have influence on the outcome compared with the Bell states corresponding to the field operator $\Psi_2(N)$. If any experiments confirmed a smaller outcome of the EPR average for maximally anti-correlated in circular basis states comparing with maximally correlated in circular basis states, it could have spoken in favor for the $N$ parameter being a finite number. On the other hand for the limit $N\to \infty$ the correlation function in reducible representation doesn't show difference with the irreducible representation except the shift phase coming from the dependence of the polarization angle on momentum. Other than that it is hard to distinguish such representation from ``standard models" for the vacuum probability density function $Z(\bm k)$ being flat in the detectors momentum solid angle spread.\\
\\
Finally the main conclusion from this work is that there may be a relativistic effect on the degree of violation in EPR-type experiments for photon fields. Two cases where considered here: where two detectors are transformed under Lorentz transformation in such a way that they still maintain in the same reference frame and where just one of the detectors is transformed under Lorentz transformation. Employing a model for relativistic EPR-type experiments in reducible representation may show the role that the vacuum probability density function $Z(\bm k)$ plays in such relativistic experiments. It turns out that assuming such not unique vacuum, the vacuum probability function $Z(\bm k)$ may have an impact on the detector outcome for such relativistic model. To see this let us schematically rewrite the results presented in previous sections. For the EPR average for two detectors $Y_\beta,Y_\alpha $ for a two-photon field operator $\Psi$ we will write
\begin{eqnarray}
&&
\langle O|
\Psi^{\dagger}
Y_\beta 
Y_\alpha 
\Psi
|O\rangle.
\end{eqnarray}
We may use the unitarity of the Lorentz transformation and assuming an invariant two-photon field operator, i.e. $U_{\Lambda}^{\dagger} \Psi U_{\Lambda}=\Psi$, we have
\begin{eqnarray}
\langle O|
\Psi^{\dagger}
Y_\beta 
Y_\alpha 
\Psi
|O\rangle
~=~
\langle O|
U_{\Lambda}U_{\Lambda}^{\dagger}
\Psi^{\dagger}
U_{\Lambda}U_{\Lambda}^{\dagger}
Y_\beta 
U_{\Lambda}U_{\Lambda}^{\dagger}
Y_\alpha 
U_{\Lambda}U_{\Lambda}^{\dagger}
\Psi
U_{\Lambda}U_{\Lambda}^{\dagger}
|O\rangle
~=~
\langle O_{\Lambda^{-1}}|
\Psi^{\dagger}
Y_{\Lambda\beta }
Y_{\Lambda\alpha }
\Psi
|O_{\Lambda^{-1}}\rangle.
\end{eqnarray}
This means that performing a transformation on the detectors $Y_{\Lambda\beta }Y_{\Lambda\alpha }$ and an compensating transformation on the vacuum field, denoted here by $O_{\Lambda^{-1}}$, would be equivalent to not performing any transformation at all.  
But when we perform a Lorentz transformation like in the first case, i.e. on both detectors in such a way that they remain in the same reference frame, the EPR average takes the form
\begin{eqnarray}
\langle O|
\Psi^{\dagger}
U_{\Lambda}^{\dagger}
Y_\beta 
Y_\alpha 
U_{\Lambda}
\Psi
|O\rangle
~=~
\langle O|
\Psi^{\dagger}
U_{\Lambda}^{\dagger}
Y_\beta 
U_{\Lambda}U_{\Lambda}^{\dagger}
Y_\alpha 
U_{\Lambda}
\Psi
|O\rangle
~=~
\langle O|
\Psi^{\dagger}
Y_{\Lambda\beta} 
Y_{\Lambda\alpha} 
\Psi
|O\rangle.
\end{eqnarray}
On the other hand we may perform an unitary transformation on states, only we have to remember that in this model we assume a not unique vacuum and invariant field operators, i.e. 
\begin{eqnarray}
&&
\langle O|
\Psi^{\dagger}
U_{\Lambda}^{\dagger}
Y_\beta 
Y_\alpha 
U_{\Lambda}
\Psi
|O\rangle
~=~
\langle O|
U_{\Lambda}^{\dagger}U_{\Lambda}
\Psi^{\dagger}
U_{\Lambda}^{\dagger}
Y_\beta 
Y_\alpha 
U_{\Lambda}
\Psi
U_{\Lambda}^{\dagger}U_{\Lambda}
|O\rangle
~=~
\langle O_{\Lambda}|
\Psi^{\dagger}
Y_\beta 
Y_\alpha 
\Psi
|O_{\Lambda}\rangle.
\end{eqnarray}
This means that performing a transformation on both detectors is equivalent to performing a transformation on the vacuum which is denoted here by $O_{\Lambda}$.
Now for the second case we perform a Lorentz transformation only on Alice's detector, i.e.
\begin{eqnarray}
\langle O|
\Psi^{\dagger}
Y_\beta
U_{\Lambda}^{\dagger} 
Y_\alpha 
U_{\Lambda}
\Psi
|O\rangle
&=&
\langle O|
\Psi^{\dagger}
Y_\beta
Y_{\Lambda\alpha} 
\Psi
|O\rangle
~=~
\langle O|
U_{\Lambda}^{\dagger} U_{\Lambda}
\Psi^{\dagger}
U_{\Lambda}^{\dagger} U_{\Lambda}
Y_\beta
U_{\Lambda}^{\dagger} 
Y_\alpha 
U_{\Lambda}
\Psi
U_{\Lambda}^{\dagger} U_{\Lambda}
|O\rangle
\nonumber\\
&=&
\langle O_{\Lambda}|
\Psi^{\dagger}
Y_{\Lambda^{-1}\beta} 
Y_{\alpha} 
\Psi
|O_{\Lambda}\rangle.
\end{eqnarray}
As we can see, performing a Lorentz transformation on Alice's detector does not have to resolve in the same EPR average as performing an inverse Lorentz transformation on Bob's detector. If any experiments confirmed such results this could have spoken in favor for a not unique vacuum representation and its impact on such relativistic experiments.
\\
\section{Acknowledgments}
I am grateful to Marek Czachor for discussions and critical comments.
\\
\appendix
\numberwithin{equation}{section}
\section{Linear and circular polarizations}
Any linear polarization for $N$-oscillator representation
can be defined due to a transformation on $a_1(\bm k,N)$
and $a_2(\bm k,N)$ annihilation operators which correspond
to linear polarizations in $x$ and $y$ direction:
\\
\begin{eqnarray}
L_{\theta}(N)^{\dagger}
a_1(\bm k,N)
L_{\theta}(N)
&=&
\cos\theta(\bm k)
a_1(\bm k,N)
+
\sin\theta(\bm k)
a_2(\bm k,N)
~=~
a_{\theta}(\bm k,N),
\nonumber\\
L_{\theta}(N)^{\dagger}
a_2(\bm k,N)
L_{\theta}(N)
&=&
-
\sin\theta(\bm k)
a_1(\bm k,N)
+
\cos\theta(\bm k)
a_2(\bm k,N)
~=~
a_{\theta'}(\bm k,N),
\end{eqnarray}
where
\begin{equation}
L_{\theta}(N)
~=~
L_{\theta}(1)^{\otimes N}
,~~~~~~~~
L_{\theta}(1)
~=~
\int d\Gamma(\bm k)
|\bm k
\rangle
\langle
\bm k |
\otimes
\exp{\left(
\theta(\bm k)(
a_1^{\dagger}
a_2
-
a_2^{\dagger}
a_1
)
\right)
}.
\end{equation}
Here $a_\theta(\bm k, N)$ and $a_{\theta'}(\bm k, N)$
hold the same commutation relation as
$a_1(\bm k, N)$ and $a_2(\bm k, N)$
\begin{eqnarray}
[a_\theta(\bm k, N),a_\theta(\bm k', N)^\dagger]
~=~
I(\bm k,N)\delta_{\Gamma}(\bm k, \bm k')
\end{eqnarray}
and $\theta'(\bm k)=\theta(\bm k)+\frac{\pi}{2}$.
One may also write this transformation in matrix form:
\begin{eqnarray}
\left(
\begin{array}{c}
a_{\theta}(\bm k,N)\\
a_{\theta'}(\bm k,N)
\end{array}
\right)
&=&
\left(
\begin{array}{cc}
\cos\theta(\bm k)&\sin\theta(\bm k)\\
\cos\theta'(\bm k)&\sin\theta'(\bm k)
\end{array}
\right)
\left(
\begin{array}{c}
a_{1}(\bm k,N)\\
a_{2}(\bm k,N)
\end{array}
\right)
\\
&=&
\left(
\begin{array}{cc}
\cos\theta(\bm k)&\sin\theta(\bm k)\\
-\sin\theta(\bm k)&\cos\theta(\bm k)
\end{array}
\right)
\left(
\begin{array}{c}
a_{1}(\bm k,N)\\
a_{2}(\bm k,N)
\end{array}
\right).
\end{eqnarray}
It should be stressed that, in the case of reducible quantization, $\theta(\bm k)$ is a function of $\bm k$ and not just a parameter. This was discussed further in section \ref{sec:trans}, where it turns out that the dependence on momentum is significant for relativistic background.\\
\\
Circular polarizations can be defined by a transformation
\begin{eqnarray}
C_{\theta}(N)^{\dagger}
a_1(\bm k,N)
C_{\theta}(N)
&=&
\frac{1}{\sqrt{2}}
\left(
a_1(\bm k,N)
+i
a_2(\bm k,N)
\right)
~=~
a_{-}(\bm k,N),
\nonumber\\
C_{\theta}(N)^{\dagger}
a_2(\bm k,N)
C_{\theta}(N)
&=&
\frac{1}{\sqrt{2}}
\left(
a_1(\bm k,N)
-i
a_2(\bm k,N)
\right)
~=~
a_{+}(\bm k,N),
\end{eqnarray}
where
\begin{equation}
C_{\theta_{ij}}(N)
~=~
C_{\theta_{ij}}(1)^{\otimes N}
,~~~~~~~~
C_{\theta_{ij}}(1)
~=~
\int d \Gamma(\bm k)
|\bm k
\rangle
\langle
\bm k |
\otimes
\exp{\left(
\sum_{ij=1,2}
\theta_{ij}~
a_i^{\dagger}
a_j
\right)
}
.
\end{equation}
Here $\theta_{ij}$ are coefficients of this transformation such that
\begin{eqnarray}
\left(
\begin{array}{cc}
\theta_{11}
&
\theta_{12}\\
\theta_{21}
&
\theta_{22}
\end{array}
\right)
~=~
\left(
\begin{array}{cc}
\frac{\sqrt{3}\pi}{9}i
-\frac{\pi}{4}i
&
-\frac{\sqrt{3}\pi}{9}
+\frac{\sqrt{3}\pi}{9}i
\\
\frac{\sqrt{3}\pi}{9}
+\frac{\sqrt{3}\pi}{9}i
&
-\frac{\sqrt{3}\pi}{9}i
-\frac{\pi}{4}i
\end{array}
\right).
\end{eqnarray}
Then the correspondences between the ladder operators in circular basis and in linear basis can be written as
\begin{eqnarray}\label{astotheta}
a_{s}(\bm k,N)
&=&
\frac{1}{\sqrt{2}}
e^{-is\theta(\bm k)}
\left(
a_{\theta}(\bm k,N)
-i s
a_{\theta'}(\bm k,N)
\right),
\\
a_{\theta}(\bm k,N)\label{athetatos}
&=&
\frac{1}{\sqrt{2}}
\sum_{s=\pm}
a_{s}(\bm k,N)
e^{is\theta(\bm k)}.
\end{eqnarray}
\\
\section{Transformation properties of null tetrad}
The spinor field transformation associated with the homogeneous Lorentz transformation
\begin{equation}
\pi_{A}(\bm{k})~~\mapsto~~ \Lambda\pi_{A}(\bm{k})
~=~
\Lambda_{A}^{~~B}\pi_{B}(\bm{\Lambda^{-1}k}).
\end{equation}
Here $\bm{\Lambda^{-1} k}$ is a space like part of a four-vector
${\Lambda^{-1}}_{a}^{~~b}k_{b}(\bm{k})$ and $\Lambda_{A}^{~~B}$ is
an unprimed SL(2,C) matrix corresponding to $\Lambda_{a}^{~~b}\in$~SO(1,3).
The null vector $k_A(\bm k)$ that plays the role of a flag-pole for the spinor field $\pi_A(\bm k)$, i.e. $k_A(\bm k)=\pi_{A}(\bm{k})\pi_{A'}(\bm{k})$, must be invariant, so
$\Lambda\pi_{A}(\bm{k})
\Lambda\pi_{A'}(\bm{k})
=\pi_{A}(\bm{k})\pi_{A'}(\bm{k})$
must be satisfied and hence
\begin{equation}
\Lambda\pi_{A}(\bm{k})
~=~
e^{-i\Theta(\Lambda,\bm{k})}
\pi_{A}(\bm{k}).
\end{equation}
The angle $\Theta(\Lambda,\bm{k})$ is called the Wigner phase. Note that in the literature it is the doubled value $2\Theta(\Lambda,\bm{k})$ which is called the Wigner phase. In analogy
\begin{equation}
\omega_{A}(\bm{k})~~\mapsto~~
\Lambda\omega_{A}(\bm{k})
~=~
\Lambda_{A}^{~~B}\omega_{B}(\bm{\Lambda^{-1}k}),
\end{equation}
and the spin-frame condition has to hold. We assume a special case, i.e.
\begin{equation}
\Lambda\omega_{A}(\bm{k})
~=~
e^{i\Theta(\Lambda,\bm{k})}
\omega_{A}(\bm{k}).
\end{equation}
It is possible to find such a spin-frame, and this was discussed in \cite{MCKW09} paper by Czachor and Wrzask. Furthermore, it is important to stress that the Wigner phase depends only on the direction of the momentum and does not depend on the frequency, so that all the parallel wave vectors correspond to the same rotational angle. This was shown, for example, by Caban and Rembieli\'nski in \cite{PCJR03}.\\\\
The null tetrad with respect to the spin-frame can be written in the form
\begin{eqnarray}
k_{a}(\bm{k})
&=&\pi_{A}(\bm{k})\pi_{A'}(\bm{k}),
\\
\omega_{a}(\bm{k})
&=&\omega_{A}(\bm{k})\omega_{A'}(\bm{k}),
\\
m_{a}(\bm{k})
&=&\omega_{A}(\bm{k})\pi_{A'}(\bm{k}),
\\
\bar{m}_{a}(\bm{k})
&=&\pi_{A}(\bm{k})\omega_{A'}(\bm{k}).
\end{eqnarray}
Here $m_{a}(\bm{k})$ and $\bar{m}_{a}(\bm{k})$ are fields associated with circular photon polarizations. Therefore their Lorentz transformation rule is
\begin{eqnarray}
m_{a}(\bm{k})~~\mapsto~~ \Lambda m_{a}(\bm{k})\label{transrule1}
~=~
\Lambda_{a}^{~~b}m_{b}(\bm{\Lambda^{-1}k})
~=~
e^{2i\Theta(\Lambda,\bm{k})}
m_{a}(\bm{k}),\\
\bar{m}_{a}(\bm{k})~~\mapsto~~ \Lambda \bar{m}_{a}(\bm{k})
~=~
\Lambda_{a}^{~~b}\bar{m}_{b}(\bm{\Lambda^{-1}k})
~=~
e^{-2i\Theta(\Lambda,\bm{k})}
\bar{m}_{a}(\bm{k}).\label{transrule1end}
\end{eqnarray}
\end{document}